\documentclass[twocolumn,amssymb, nobibnotes, aps, prx]{revtex4-2}
\usepackage{amsmath}
\usepackage{amssymb}
\usepackage{graphicx}
\usepackage{tikz}
\usepackage{float}
\usepackage{soul}

\begin{document}

\title{Flat bands and entanglement in the Kitaev ladder}
\author{Ritu Nehra}
\email{ritunehra@iiserb.ac.in}
\author{Devendra Singh Bhakuni}
\email{devendra123@iiserb.ac.in}
\author{Ajith Ramachandran}
\email{ajithr@iiserb.ac.in}
\author{Auditya Sharma}
\email{auditya@iiserb.ac.in}
\affiliation{Department of Physics, Indian Institute of Science Education and Research, Bhopal, India}
\begin{abstract}
  We report the existence of \emph{flat bands} in a p-wave
  superconducting Kitaev ladder. We identify two sets of parameters
  for which the Kitaev ladder sustains flat bands. These flat bands
  are accompanied by highly localized eigenstates known as compact
  localized states. Invoking a Bogoliubov transformation, the Kitaev
  ladder can be mapped into an interlinked cross-stitch lattice. The
  mapping helps to reveal the compactness of the eigenstates each of
  which covers only two unit cells of the interlinked cross-stitch
  lattice. The Kitaev Hamiltonian undergoes a topological-to-trivial
  phase transition when certain parameters are fine-tuned. Correlation
  matrix techniques allow us to compute entanglement entropy of the
  many-body eigenstates. The study of entanglement entropy affords
  fresh insight into the topological phase transitions in the
  model. Sharp features in entanglement entropy when bands cross
  indicate a deep underlying relationship between entanglement entropy
  and dispersion.
\end{abstract}

\maketitle

\section{Introduction}
   In the last few years, dispersionless bands, also known as flat
   bands have received a great deal of attention within the condensed
   matter physics
   community~\cite{parameswaran2013fractional,flach2014detangling,leykam2018artificial}.
   In tight binding systems which have periodic lattice potentials, the
   Bloch theorem ~\citep{ashcroft2010solid} assures the existence of a well defined band
   structure for a single particle, which in turn, uniquely defines
   the group velocity of the particle.  Certain carefully tuned
   lattice Hamiltonians may exhibit one or more \emph{flat bands} in
   the band structure, where the energy is independent of momentum
   resulting in zero group velocity.  The study of
   the suppression of wave transport induced by flat bands has been carried out in a number of works~\cite{mielke1991ferromagnetism,tasaki1992ferromagnetism,flach2014detangling}.
   Since the wave equations which govern the dynamics can be generalised, flat band physics has been explored
   in diverse systems ranging from Hubbard models~\cite{mielke1999ferromagnetism,tasaki1998nagaoka} to photonic crystals~\cite{guzman2014experimental,vicencio2015observation,mukherjee2015observation} and
   Bose-Einstein condensates~\cite{zhang2013bose, gladchenko2009superconducting}.
   
   The eigenstates corresponding to flat bands are highly localized,
   covering a finite number of lattice sites, and are therefore referred to as
   compact localized states (CLS)~\cite{flach2014detangling}.  One of
   the prime strategies for finding flat band systems has been to
   first identify CLS, and then engineer Hamiltonians that could
   support them.  A number of recent
   studies~\cite{mielke1991ferromagnetism,maimaiti2017compact,ramachandran2017chiral}
   have searched for flat band generating algorithms with the aim of
   identifying more and more flat band networks. Many such studies
   have been based on non-interacting particles, and focus on
   geometries which support a flat band through destructive
   interference properties of the eigenstates
   ~\cite{morales2016simple,maimaiti2017compact,maimaiti2019universal}. In
   the present work, we report the existence of flat bands in a rather
   simple superconducting Kitaev ladder Hamiltonian system, which has
   hitherto been unexplored in this context. Subjecting the Kitaev
   ladder to a Bogoliubov transformation allows us to map it to an
   interlinked cross-stitch lattice and thus obtain the explicit
   compact localized states in the transformed basis. One key novelty
   in our work is that despite the very simple geometry of the
   lattice, a flat band becomes possible due to the p-wave
   superconducting term. Our tests indicate that the introduction of a
   pairing term to generate flat bands may well be quite general
   beyond the specifics of the current model, and could potentially
   open up a new avenue for flat-band engineering.
\begin{figure}[h!]
\begin{center}
\includegraphics[ ]{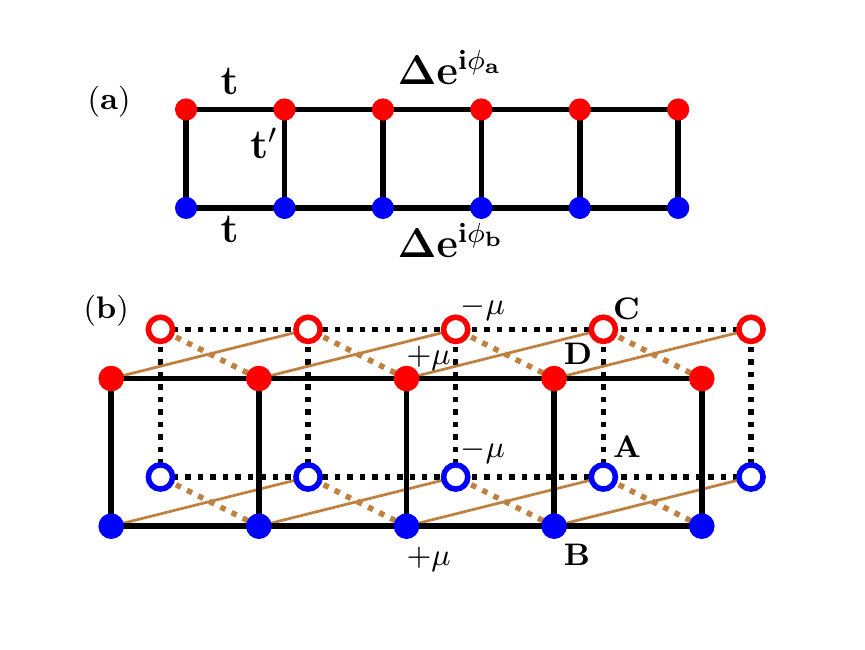}
\caption{ Schematic representations of (a) the Kitaev
  ladder, and (b) Interlinked cross-stitch lattice. The solid (dotted)
  lines show positive (negative) hopping. The open and closed circles
  represent negative and positive onsite energies
  respectively. Using a Bogoliubov transformation,
    the Kitaev chain Hamiltonian can be mapped to the interlinked
    cross-stitch lattice Hamiltonian.}
\label{mapping}
\end{center}
\end{figure}

   The Kitaev ladder~\citep{Maiellaro2018} is a generalization of the
   Kitaev chain~\citep{kitaev2001unpaired}, which is a canonical
   system that admits a topological phase
   transition~\citep{wu2012topological}. The simultaneous presence of
   both the superconducting term and the ladder geometry leads to many
   rich phenomena including some exotic transport
   properties~\citep{nehra2019enhancement}.  Furthermore, this system
   exhibits multiple topologically non-trivial phases characterized by
   different symmetry classes, which may be accessed by tuning
   different parameters. Quantum
   entanglement~\citep{flammia2009topological,tasaki1992ferromagnetism,PhysRevB.98.045120,chen2013quantum,RevModPhys.82.277}
   has emerged as a key tool of study whenever quantum phase
   transitions are present. This includes topological-to-trivial phase
   transistions~\citep{kitaev2006topological,PhysRevLett.113.156402,liu2016topological},
   Meissner-to-vortex phase
   transistions~\citep{nehra2018many,hugel2014chiral} ,
   localization/de-localization
   transitions~\citep{roy2018entanglement,roy2019study} and many
   others~\cite{dey2019quantum,sharma2015landauer,sable2018landauer,PhysRevB.98.045408}. In
   this paper, we study entanglement in the many-body eigenstates of
   the Kitaev ladder. Since the computation of entanglement in the
   eigenstates of systems that feature a superconducting term is
   challenging, the literature~\cite{vitagliano2010volume} on this
   subject is rather sparse, and our work is to be seen as a
   contribution to filling this void. Thus, our paper is a study of the
   interplay of flat bands, topological properties, and entanglement
   properties of the Kitaev ladder. 

  The layout of the article is as follows: we start by introducing
   the Hamiltonian for the two-leg Kitaev ladder. The conditions for
   the flat bands and a discussion of compact localized states are
   provided in the subsequent section. The next section discusses the
   topological properties of the system before moving on to a study of
   entanglement entropy.  Finally, we summarize our main findings in
   the last section.

\section{Kitaev ladder}
The Kitaev ladder~\citep{kitaev2001unpaired,nehra2019transport} (Fig.~\ref{mapping}(a)) consists of two Kitaev chains connected to one another through inter-leg hopping. This system is described by the tight-binding Hamiltonian:
\begin{equation}
\begin{split}
H=&-t\displaystyle\sum_{n\atop \sigma=1,2}c_{n+1,\sigma}^{\dagger}c_{n,\sigma}-\mu\displaystyle\sum_{n\atop \sigma=1,2}c_{n,\sigma}^{\dagger}c_{n,\sigma}-t^{\prime}\displaystyle\sum_{n}c_{n,1}^{\dagger}c_{n,2}\\&-\Delta\displaystyle\sum_{n\atop \sigma=1,2}e^{i\phi_{\sigma}}c_{n+1,\sigma}^{\dagger}c_{n,\sigma}^{\dagger}+H~.~c~.
\end{split}\label{Ham_rs}
\end{equation}
where, $t$ is the intra-leg hopping amplitude,
$c^{\dagger}_{n,\sigma}$($c_{n,\sigma}$) are creation (annihilation)
operators on the $n^{th}$ site of the ladder with $\sigma=1,2$ running
over two legs of the ladder, and $\mu$ is the on-site chemical
potential. The inter-leg hopping on the ladder is $t^{\prime}$ and the
superconducting gap is $\Delta$ with a phase factor
$e^{i\phi_{\sigma}}$ in each leg of the ladder. As the annihilation of
an electron is equivalent to the creation of a hole, one can write
$d^\dagger_{n,\sigma}=c_{n,\sigma},d_{n,\sigma}=c^\dagger_{n,\sigma}$,
and the Hamiltonian can be redefined in momentum space as
~\citep{nehra2019enhancement,nehra2019transport},
\begin{align}
H=\displaystyle\sum_{k}{\Gamma}^{\dagger}_{k}\mathcal{H}(k){\Gamma}_{k}
\end{align}
where
\begin{align}
{\Gamma}^{\dagger}_{k}=\begin{bmatrix}
c_{k,1}^{\dagger}& d_{k,1}^{\dagger}& c_{k,2}^{\dagger}& d_{k,2}^{\dagger}
\end{bmatrix},\;
{\Gamma}_{k}=
\begin{bmatrix}
c_{k,1}\\ d_{k,1}\\ c_{k,2}\\ d_{k,2}
\end{bmatrix}.
\end{align}
Thus the overall Hamiltonian in $k-$space can be given as,
\begin{align}\small
\mathcal{H}(k)=\begin{bmatrix}
-\epsilon_{\mu ,k} &t_{\Delta,\phi_1} &-t^{\prime}& 0\\
t^*_{\Delta,\phi_1} & \epsilon_{\mu ,k} & 0&t^{\prime}\\
-t^{\prime}& 0&-\epsilon_{\mu ,k} &t_{\Delta,\phi_2} \\
 0&t^{\prime}&  t^*_{\Delta,\phi_2} & \epsilon_{\mu ,k} 
\end{bmatrix} \label{ham_k}
\end{align}
with $\epsilon_{\mu ,k} = (2t\cos{k}+\mu)$ and $t_{\Delta,\phi_j}=-2i\Delta\sin{k}e^{i\phi_{j}}$ ; $j=1,2$ and $^*$ denotes the complex
conjugation. Diagonalizing the Hamiltonian, the dispersion relation is given by
\begin{equation}
 E(k)=\pm\sqrt{\epsilon_{\mu ,k}^2+{t^{\prime}}^2+\tau_{\Delta,k}\pm 2t^{\prime}\sqrt{\epsilon_{\mu ,k}^2+\tau_{\Delta,k}{\sin}^2 \frac{\Phi}{2}}}\label{disp}
\end{equation}
where, $\tau_{\Delta,k} = 4{\Delta}^2 {\sin}^2 k$ and
$\Phi=\phi_2-\phi_1$.  The ladder supports four bands in the band structure due to bonding and antibonding of both holes and electrons. 

\subsection{Flat band and localization}
Flat bands are dispersionless bands in the energy spectrum of the
translationally invariant lattice Hamiltonian. The Hamiltonian
is fine-tuned to specific parameter values, leading to strictly flat
bands in the entire Brillouin zone. A network could possess one or more
flat bands that may be isolated or involve crossings with
dispersive bands. In all the known cases, eigenstates corresponding to
isolated flat bands are compact localized eigenstates (CLS) which reside on a finite volume of the lattice and absolutely vanishing in the rest
of the lattice \cite{maimaiti2019universal}. 
\begin{figure}[ht!]
\includegraphics[scale=0.33]{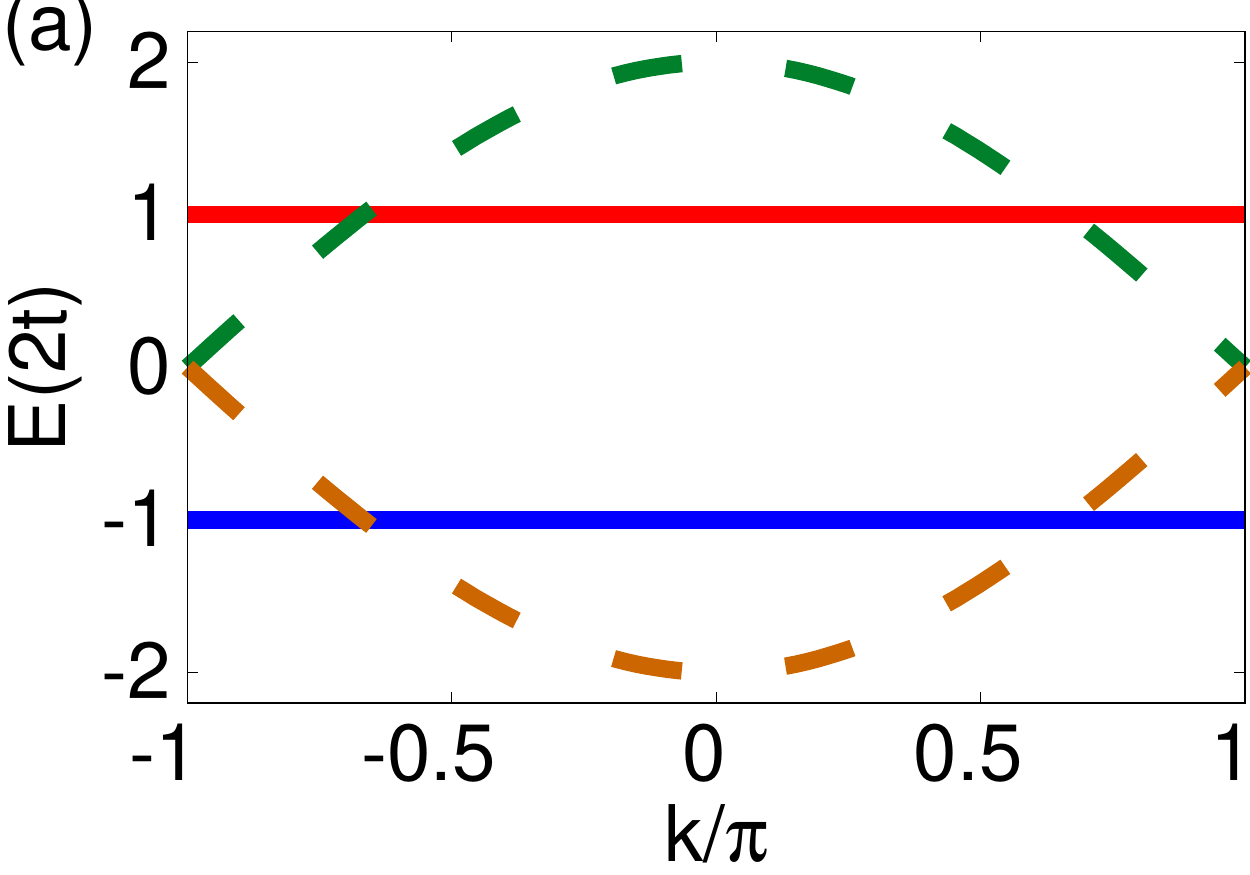}
\includegraphics[scale=0.33]{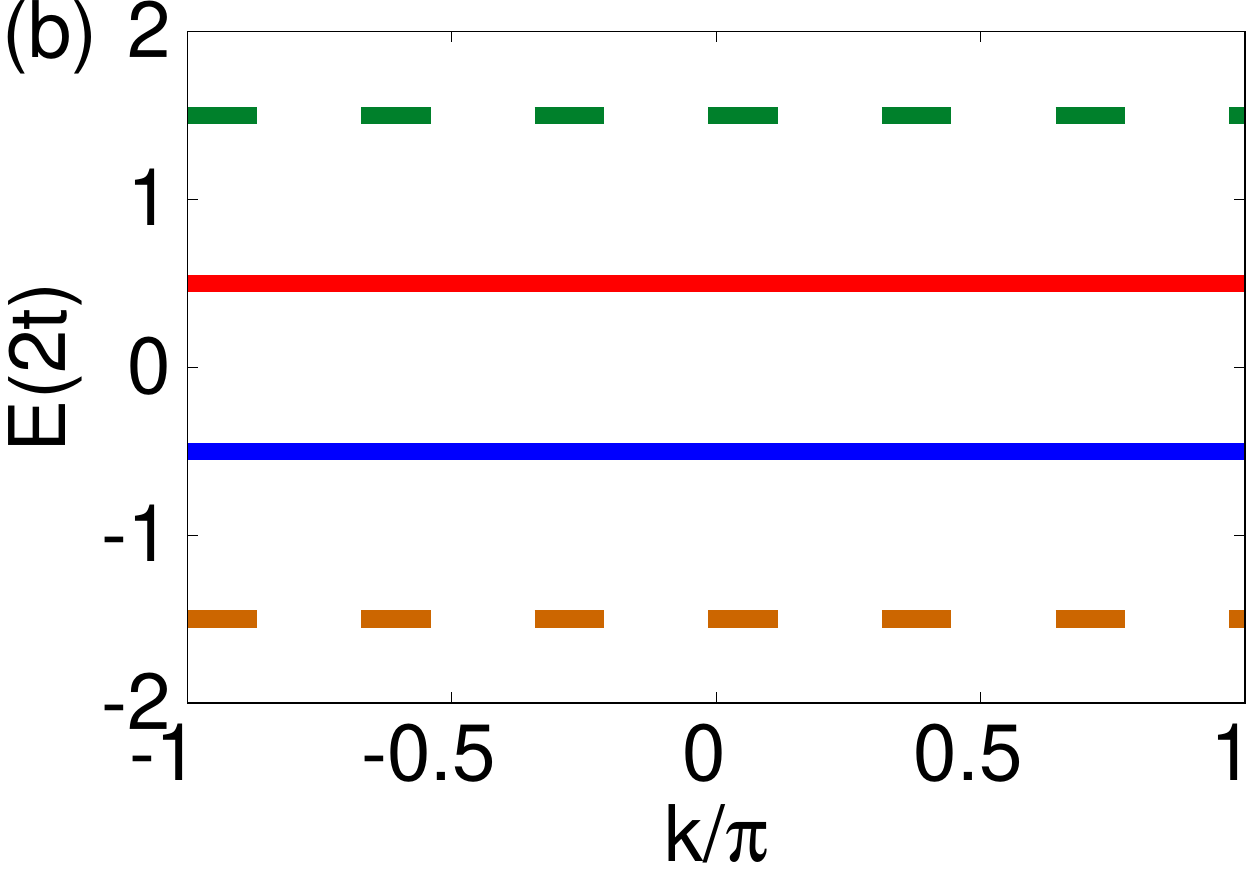}
\caption{The dispersion spectrum of system with (a) $\Phi = 0,~\mu = t' =1,~ \Delta = t=1$, and (b)$\Phi = \pi,~\mu = 0,~ \Delta = t=1,t'=1$.}
\label{dispersion}
\end{figure}
In some particular cases,
when flat bands make an intersection with a dispersive band, a new
type of extended eigenstate usually called as {\it line state} exists~\cite{xia2018unconventional}.

The ladder shown in Fig.~\ref{mapping} (a) without the p-wave pairing
term (i.e., $\Delta = 0$) possesses two dispersive bands. The
introduction of the p-wave pairing term, along with a careful choice
of the parameters can help to engineer dispersionless bands. There are
two main parameter sets that we consider in this study: 1) $\Phi =
0,~\mu = t',~ \Delta = t$ and 2) $\Phi = \pi,~\mu = 0,~ \Delta =
t$. The first parameter set results in a band structure with two
dispersive and two flat bands while the second results in four flat
bands (Fig.~\ref{dispersion}).  There maybe other parameter sets which
yield flat bands, and perhaps they maybe found with the help of a
recently proposed prescription~\cite{toikka2018necessary}.  We point
out that the Kitaev chain itself admits flat bands - the flat band
condition and the corresponding CLS here are discussed in the
Appendix.

To obtain a finer understanding of the flat band conditions, it is useful to rewrite the Hamiltonian
in Eq. \ref{ham_k} as:
\begin{widetext}
\begin{align}
\mathcal{H}(k)=\begin{bmatrix}
-te^{ik}-te^{-ik}-\mu & -\Delta e^{ik}e^{i\phi_{1}} + \Delta e^{-ik}e^{i\phi_{1}} & -t^{\prime} & 0\\
\Delta e^{ik}e^{-i\phi_{1}}-\Delta e^{-ik}e^{-i\phi_{1}} &  te^{ik}+te^{-ik}+\mu & 0 & t^{\prime}\\
-t^{\prime} & 0 & -te^{ik}-te^{-ik}-\mu & -\Delta e^{ik}e^{i\phi_{2}}+\Delta e^{-ik}e^{i\phi_{2}}\\
 0 & t^{\prime} &  \Delta e^{ik}e^{-i\phi_{2}}-\Delta e^{-ik}e^{-i\phi_{2}} & te^{ik}+te^{-ik}+\mu.
\end{bmatrix}
\label{ham_k_iCS}
\end{align}
\end{widetext}
In this form, the Hamiltonian maybe interpreted to represent a new lattice as shown in Fig.~\ref{mapping}(b). 
Equations
\ref{ham_k} and \ref{ham_k_iCS} are equivalent, with
Eq. \ref{ham_k_iCS} representing the hopping configurations of the
electron and hole explicitly.  The new lattice, an interlinked
cross-stitch lattice (ICS) consists of a unit cell with four sites
which repeats in one spacial direction generating the whole
lattice. The mapping leaves the dispersion relation (Eq. \ref{ham_k})
unchanged resulting in the same four bands in the dispersion diagram.
The ICS Hamiltonian also possesses the particle-hole
symmetry. The ICS is a modified version of the well-studied
cross-stitch lattice, which under certain conditions supports flat
bands.
\subsubsection{$\Phi = 0,~\mu = t',~ \Delta = t$}  
The parameter set results in the dispersion relation $E = \pm 2 t,
\pm2\sqrt{t^2+t'^2+2tt'\cos[k]}$ which corresponds to two flat bands and two dispersive
bands as shown in Fig.~\ref{dispersion} (a). Keeping the constraints
$~\mu = t',~ \Delta = t$ and varying the parameters, the flat bands
can be positioned to cross the dispersive bands or kept isolated from
the dispersive bands. This choice of parameters brings forth some
interesting topological properties which are discussed in the next
section. 
\begin{figure}[ht!]
  (a)\includegraphics[scale=0.82]{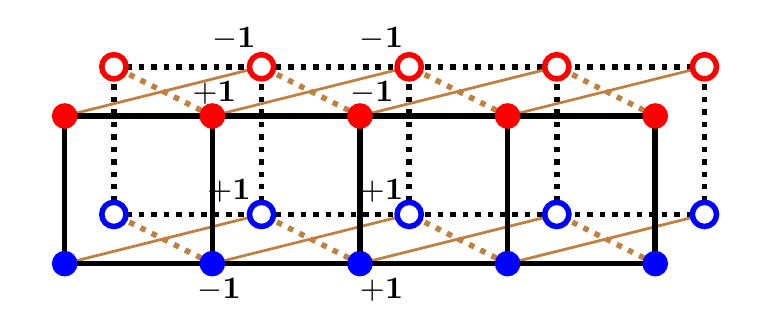}
  (b)\includegraphics[scale=0.82]{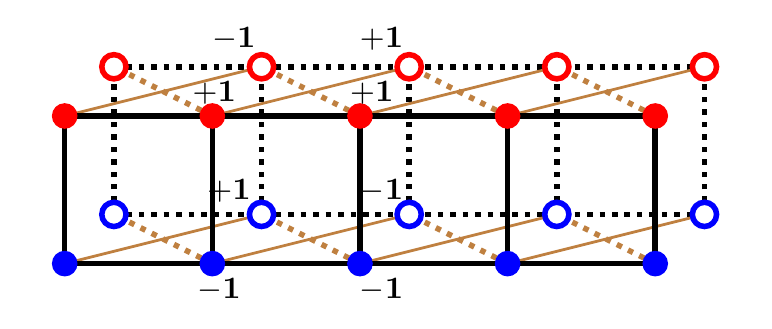}
  \caption{The compact localized states corresponding to (a) $E=-2$, (b) $E=2$ and system parameters are $\Phi = 0,~\mu = t' = 1,~ \Delta = t =1$. The dotted lines represent hopping $-1$ and regular lines represents hopping $+1$. The CLS resides on eight lattice sites with amplitude $\pm 1$ on those sites.}
  \label{cls_0}
\end{figure}
The CLS corresponding to each flat band can be identified
as shown in Fig.~\ref{cls_0}. The CLS resides on two unit cells,
i.e., eight lattice sites, and has strictly zero amplitude outside
the two unit cells which is guaranteed by the destructive interference
of the probability amplitudes of the CLS at these sites. A translation of relevant unit cell across
the length of the lattice results in the other CLS. Thus, the number of
CLS for each flat band depends on the system size. 

\subsubsection{$\Phi = \pi,~\mu = 0,~ \Delta = t$}  
This parameter set results in the dispersion relation $E=\pm(2t\pm
t')$ yielding four flat bands as shown in Fig.~\ref{dispersion}(b).
\begin{figure}[ht!]
  (a)\includegraphics[scale=0.82]{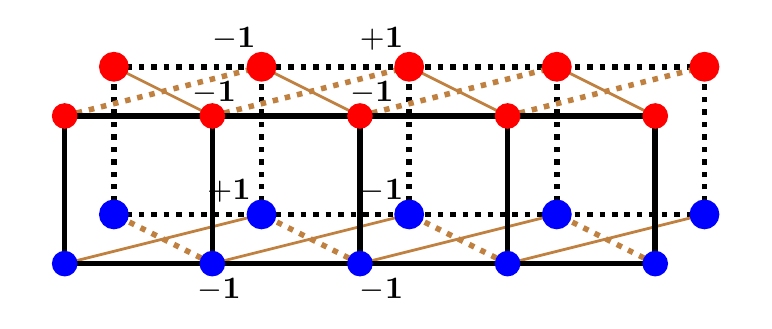}
  (b)\includegraphics[scale=0.82]{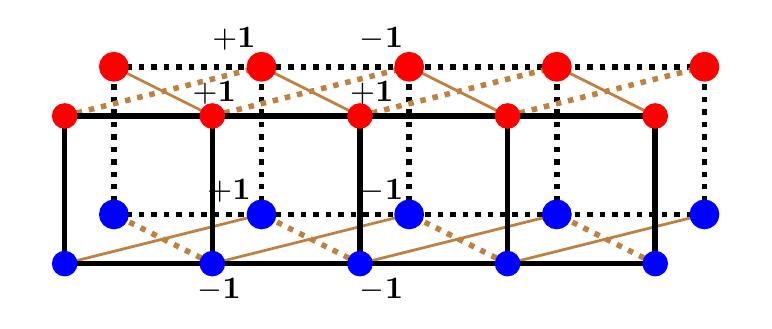}
  (c)\includegraphics[scale=0.82]{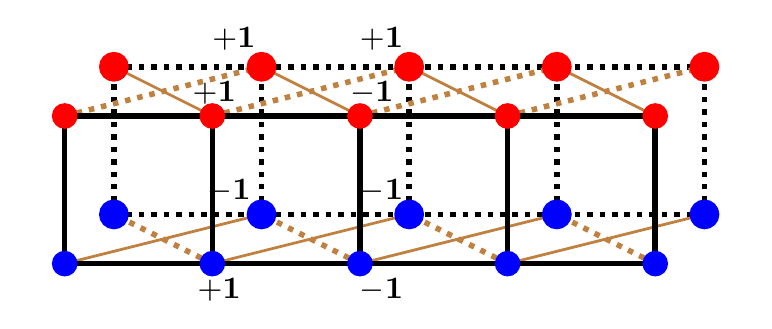}
  (d)\includegraphics[scale=0.82]{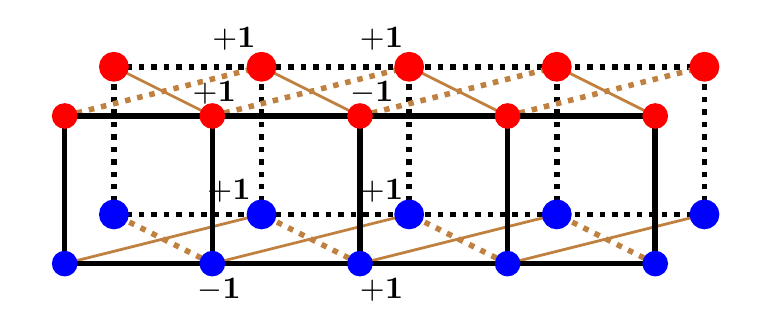}
  \caption{The compact localized states coresponding to (a) $E=3$, (b) $E=1$, (c) $E=-1$, (d) $E=-3$ and system parameters are $t=\Delta=1$, $\mu=0$, $t'=1$, $\phi_1=0$, $\phi_2=\pi$. The dotted lines represent hopping $-1$ and regular lines represents hopping $+1$. The CLS resides on eight lattice sites with amplitude $\pm 1$ on those sites. }
  \label{cls}
\end{figure}
The eigenstates corresponding to the flat bands are the compact
localized states (CLS) which are identified in Fig.~\ref{cls}. 
This CLS also resides on two unit cells, i.e., eight lattice sites, and has
strictly zero amplitude outside the two unit cells as before. Thus,
the eigenstate corresponding to the the original Kitaev ladder would
also be localized and the transport will be prohibited for these
parameter sets leaving the ladder in an insulating phase.
Furthermore, the condition $t' = 0$ leads to two uncoupled regular
cross-stitch chains, each possessing two flat bands. 

\subsection{Toplological properties}
One of the most striking features of the Kitaev
ladder~\citep{kitaev2001unpaired,PhysRevB.95.195160} is its
topological property. The Kitaev ladder encounters a change of
topological class~\citep{kitaev2009periodic,wu2012topological} from
BDI to D ~\citep{Maiellaro2018,PhysRevB.99.235112} when subjected to
a superconducting phase that breaks time-reversal symmetry. The BDI
class Kitaev ladder undergoes a quantum phase transition when
the winding number~\citep{Maiellaro2018,PhysRevB.95.195160} changes from
two to one to zero in the absence of superconducting
phase~\citep{nehra2019transport}.
\begin{figure}[h!]
\includegraphics[scale=0.336]{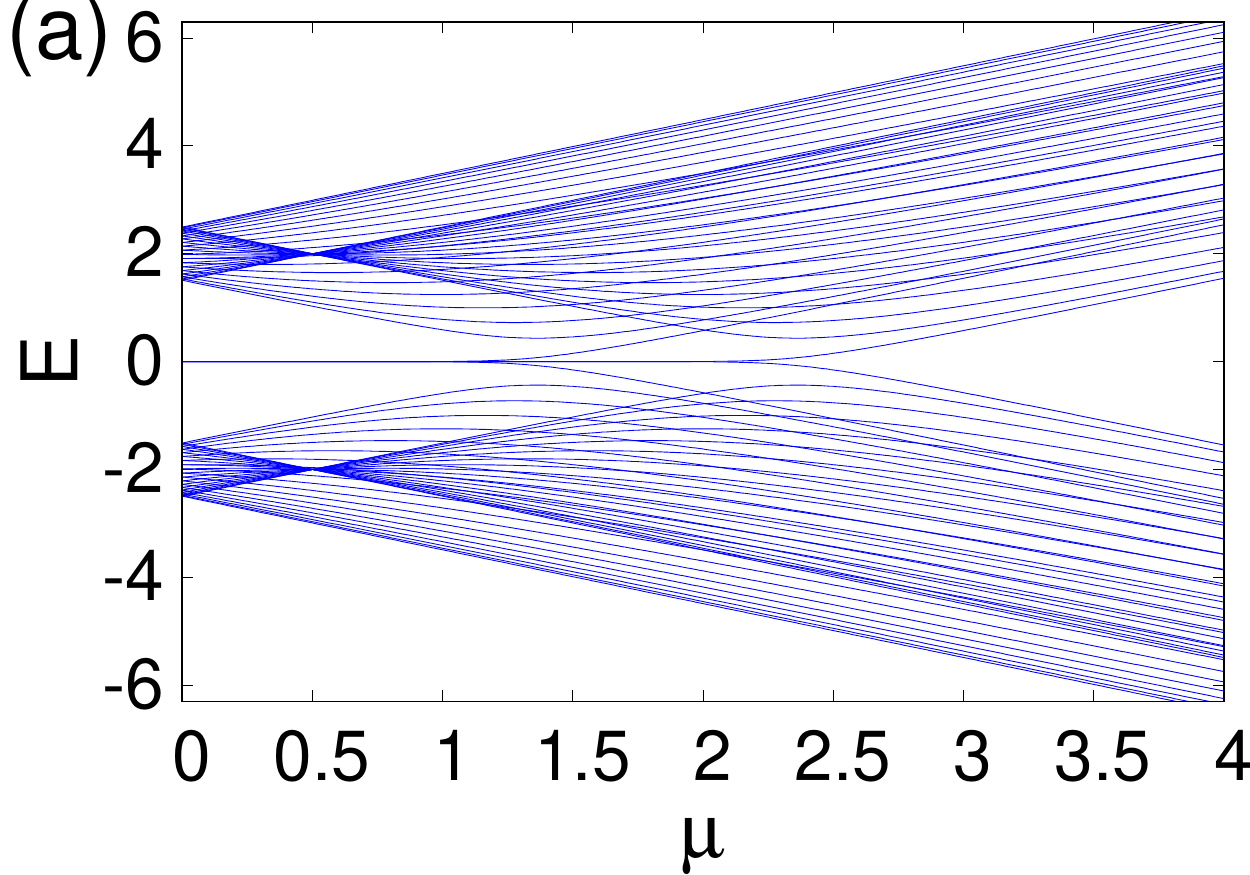}
\includegraphics[scale=0.336]{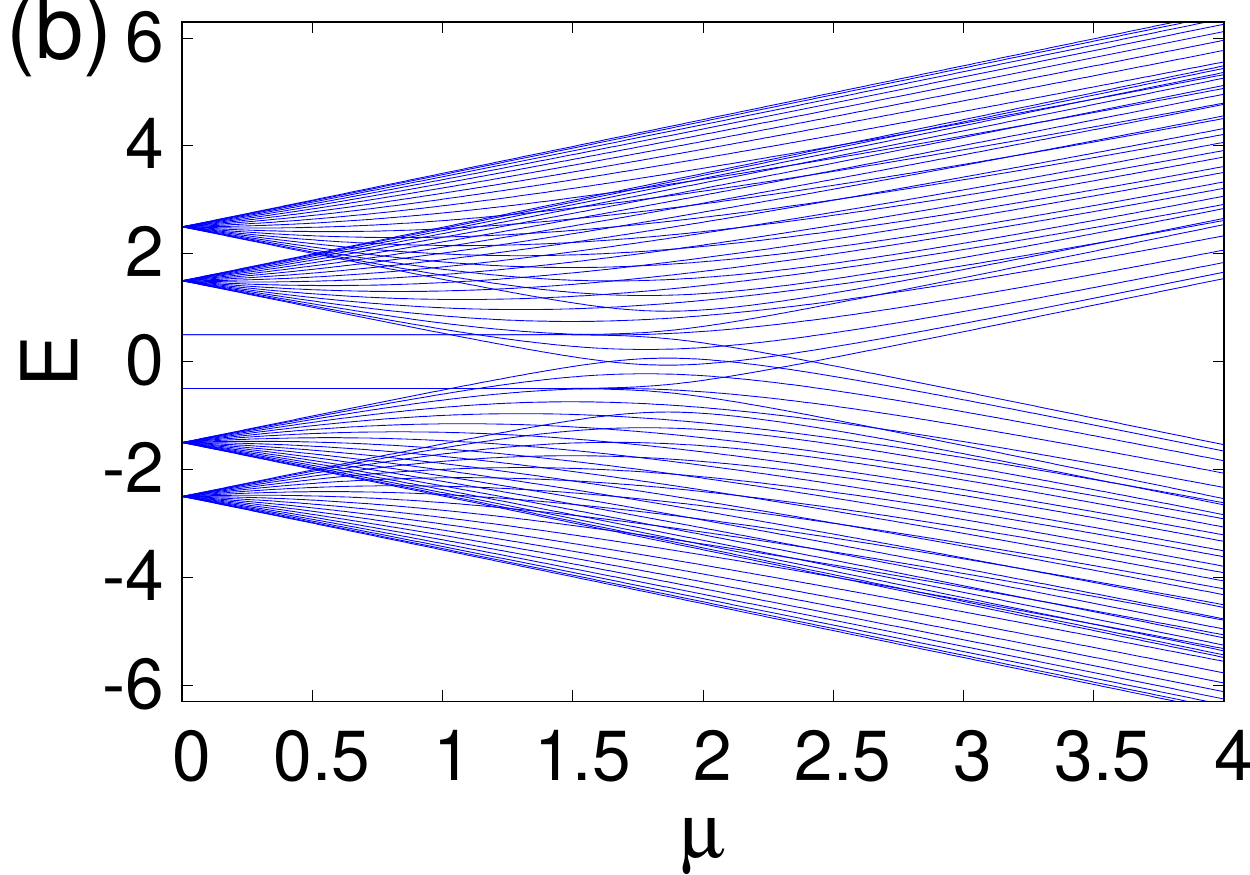}
\caption{The single particle energy spectrum of kitaev ladder with various on-site chemical potential $\mu$ under imposed open boundary conditions. The other system parameters are choosen to be $L=20$, $t=\Delta=1.0$, $t'=0.5$ with (a) $\Phi=0$, (b) $\Phi=\pi$.}
\label{fig:4}
\end{figure}
The energy spectrum in the absence of superconducting phase difference
$\Phi$ is shown in Fig.~\ref{fig:4}(a). The topological region is of
two types. The region $|\mu|< 2t-t'$, is characterized by the presence
of four zero energy Majorana edge states with winding number two,
while the region $2t-t' < |\mu| < 2t+t' $ is characterized by the
presence of two zero energy Majorana edge states with winding number
one. The trivial phase with winding number zero lies beyond
$|\mu|=2t+t'$ featuring no zero energy Majorana states.
Figure~\ref{fig:4}(b) shows the energy spectrum when the superconducting
phase difference is set to be $\Phi = \pi$. We see two distinct
two-fold degenerate states at $\pm \frac{t'}{t}$ indicating a partial
lifting of the four-fold degeneracy of the zero energy states in
Fig.~\ref{fig:4}(a). Also, the system is now classified into the D
class, because of the broken time-reversal symmetry. The
topological-to-trivial phase transition is also signalled by the
Majorana number $M$ ~\citep{WU20123530,nehra2019transport}. We see
multiple crossings of energy bands starting from $|\mu|=2t-t'$ to
$|\mu|=2t+t'$. These crossings of bands increase the
degeneracies of the system and are nicely captured by entanglement
entropy as shown in the next section.

\section{Entanglement Entropy}
Entanglement~\citep{flammia2009topological,PhysRevB.98.045120} is one
of the most fundamental aspect of many body quantum systems. It
reveals the quantum correlations between the different partitions
under study. The most commonly used quantifier of
entanglement~\citep{Peschel_2009} for a pure state is the bipartite
von Neumann entropy : $S_{A}=-\text{Tr}(\rho_{A}\text{ln}\rho_{A})$,
with $\rho_{A}$ being the reduced density matrix of the subsystem $A$
calculated by tracing out the degrees of freedom of the subsystem
$B$. In the non-interacting limit with no pairing term, the
entanglement entropy of a subsystem can be effectively calculated
with the help of the subsystem correlation matrix:
$C_{mn}=\langle c_{m}^{\dagger}c_n\rangle$. It has been
shown~\citep{peschel2003calculation,peschel2012special,PhysRevB.98.045120}
that the entanglement entropy is connected to the eigenvalues of this
correlation matrix. The correlation matrix approach can be further
generalized to study Hamiltonians including a pairing term by considering an
additional correlation matrix: $F_{mn}=\langle
c_{m}^{\dagger}c_{n}^{\dagger}\rangle$. 

To study this system, we use the well known LSM
method~\citep{lieb1961two} introduced by Lieb, Schultz and
Mattis. Using the LSM formalism, one can write the Hamiltonian in
Eq. (\ref{Ham_rs}) as the sum of a Hermitian matrix $A$ and a
non-Hermitian matrix $B$ as given in Eq. (\ref{eq:ab}):
\begin{equation}
\label{eq:ab}H=\sum_{i,j=1}^{N} c_{i}^{\dagger}A_{ij}c_{j} + \frac{1}{2}\sum_{i,j=1}^{N}(c_{i}^{\dagger}B_{ij}c_{j}^{\dagger}-c_{i}B^{*}_{ij}c_{j}),
\end{equation}  
where, $c_i$ is a fermionic annhilation operator on the $i^{th}$ site
of the leg $a$ for $i=1,2,\hdots,L$ and on the $(i-L)^{th}$ site of
the leg $b$ for rest of the values of $i$, $N=2L$ being the total number of
sites of the ladder system.  Upon diagonalization with the help of the Bogoliubov transformation
$\eta_{\alpha}=\sum_{i=1}^{N}\left(\frac{1}{2}\left(\Phi_{\alpha}(i)+\Psi_{\alpha}(i)\right)c_i+\frac{1}{2}\left(\Phi_{\alpha}(i)-\Psi_{\alpha}(i)\right)c_{i}^{\dagger}\right)$, the Hamiltonian takes the form:
\begin{equation}
H=\sum_{\alpha=1}^{N}\Lambda_{\alpha}\left(\eta_{\alpha}^{\dagger}\eta_{\alpha}-1/2\right).
\end{equation} 
The normalized vectors $\Phi_{\alpha}$, $\Psi_{\alpha}$ and the energy
spectrum $\Lambda_{\alpha}$ can be obtained by solving two coupled
equations ~\citep{Mahyaeh_2018} for $\Phi_{\alpha}$ and
$\Psi_{\alpha}$:
\begin{eqnarray}
\left(A-B\right)\Phi_{\alpha}=\Lambda_{\alpha}\Psi_{\alpha}\\
\left(A+B\right)\Psi_{\alpha}=\Lambda_{\alpha}\Phi_{\alpha}
\end{eqnarray}

In order to compute~\citep{PhysRevB.91.220101} the entanglement
entropy, one has to calculate the correlation matrix of
the system. The full system correlation matrix is given by
\begin{equation}
G_{ij}=\langle GS|(c_{i}^{\dagger}-c_{j})(c_{i}^{\dagger}+c_{j})|GS\rangle
\end{equation}
where $i,j = 1, 2, \hdots , N$.
This correlation matrix can be computed using normalized functions $\Phi_{\alpha}$ and $\Psi_{\alpha}$ as
\begin{equation}
G_{ij}=-\sum_{\alpha=1}^{N}\Psi_{\alpha}(i)\Phi_{\alpha}(j).
\end{equation}
The anticommutation properties of the $\eta$ operators allow us to
obtain the useful relations:
$\langle\eta^{\dagger}_i\eta_j\rangle=\delta_{ij}$ and
$\langle\eta_i\eta_j\rangle=0$.  To compute von Neumann entanglement
entropy, one first selectively takes out the part of the full
correlation matrix which connects to the subsystem of interest as
$G_s$. Then the entanglement entropy is computed using the method
discussed in Ref.~\citep{vitagliano2010volume}. Thus,
\begin{equation}
S=-\left(\frac{1+\nu_{\alpha}}{2}\right)\text{ln}\left(\frac{1+\nu_{\alpha}}{2}\right)-\left(\frac{1-\nu_{\alpha}}{2}\right)\text{ln}\left(\frac{1-\nu_{\alpha}}{2}\right)
\end{equation} 
where, $\nu_{\alpha}$ is the square root of the eigenvalues of the
matrix $G_sG^{T}_s$.

In this work, we analyze two sets of parameters that yield flat bands,
and employ entanglement entropy to study the similarities and
differences of their properties. Furthermore, in each of these cases,
it is useful to consider two types of partitions of the system, which
we call horizontal and vertical divisions. The horizontal division
makes the two legs of the ladder as two partitions of the system
whereas the vertical division considers the two halves of the ladder
with ladder length $\frac{L}{2}$ as the two subsystems of interest.

\subsection{Phase difference \textbf{$\Phi=0$}}
\begin{figure}[h!]
\includegraphics[scale=0.3366]{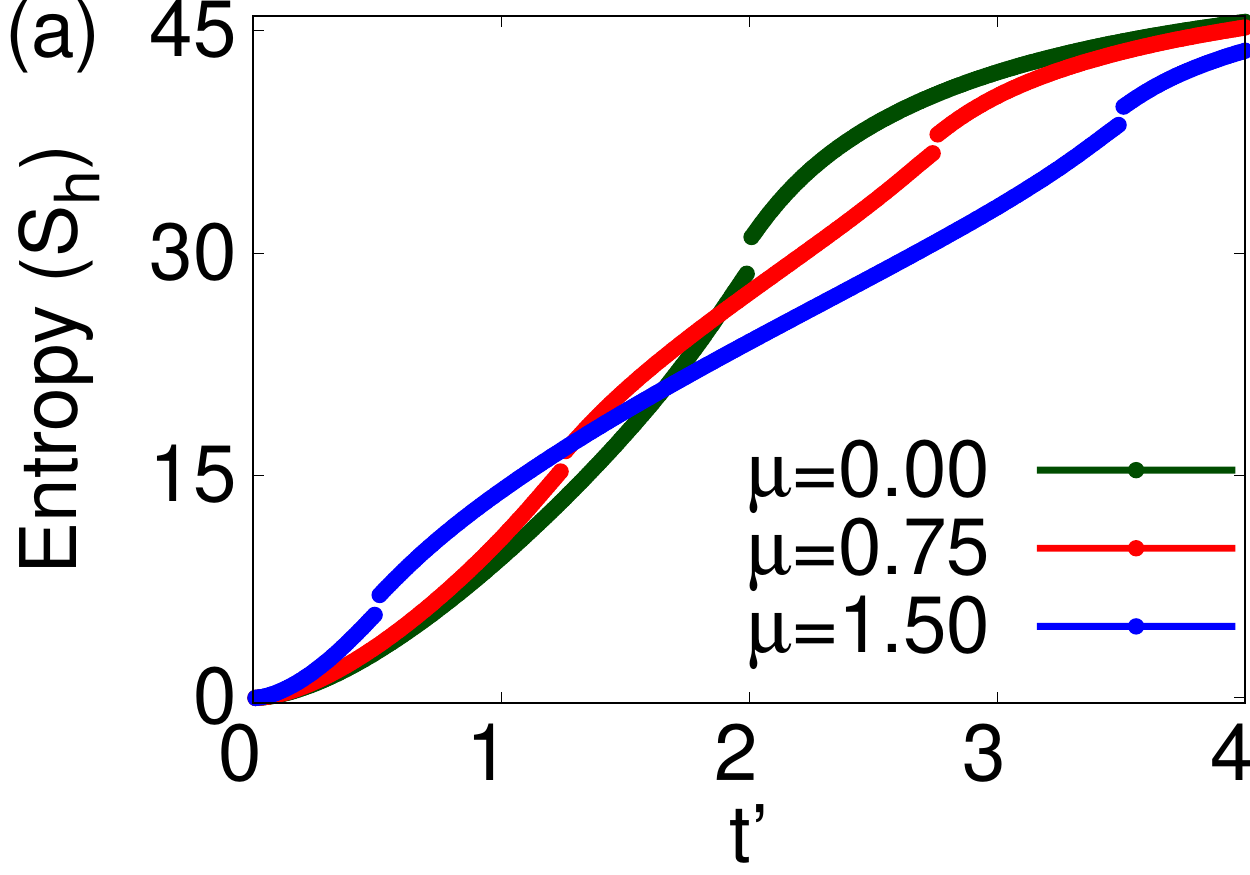}
\includegraphics[scale=0.3366]{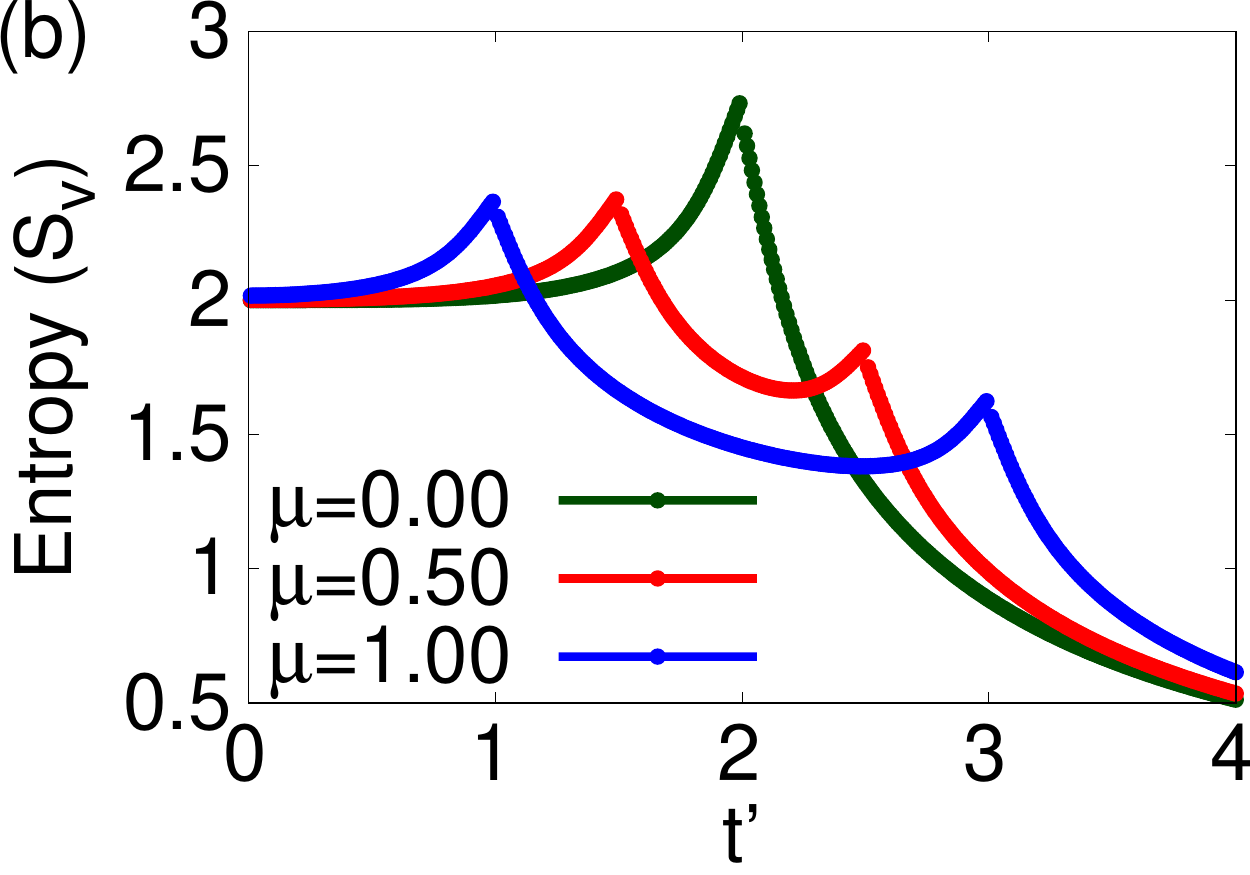}
\caption{(a) The entanglement entropy ($S_h$) of leg $a$ of the Kitaev
  ladder with respect to the other leg $b$ in units of $\ln 2$. (b)
  The entanglement entropy ($S_v$) of half the Kitaev ladder with
  respect to the other half of the ladder in units of $\ln 2$ for the
  case of vertical division. The parameters are chosen to be
  $L=50$, $t=\Delta=1.0$, $\Phi=0\;(\phi_1=\phi_2=0)$ with various $t'$ along with
  different $\mu$ values under periodic boundary conditions.}
\label{fig:5}
\end{figure}

In this subsection, the superconducting phase difference between the
two legs of the ladder is maintained at $\Phi=0\;(\phi_1=\phi_2=0)$. 
The dispersion relation of the Kitaev ladder system under this condition is (shown in Fig.~\ref{fig:4}(a))
\begin{equation}\label{eq14}
E(k)=\pm\sqrt{(2t\cos k+\mu\pm t')^2+4\Delta^2\sin^2 k}.
\end{equation}
The inter-leg hopping $t'$ results in intermixing of two Kitaev chain
bands, whose effects are well studied with the help of entanglement
entropy.  Firstly, the behaviour of entanglement entropy for various
inter-leg hopping $t'$ of the Kitaev ladder is shown for the two type
of partitions in Fig.~\ref{fig:5}.
As depicted in Fig.~\ref{fig:5}(a) in the absence of inter-leg hopping
$t'$, the two legs are disconnected which results in zero entanglement
entropy between both legs. As inter-leg hopping $t'$ is increased, the
two legs start to connect through rungs hopping which contributes
towards the entanglement entropy between the two legs. The system
exhibits two quantum phase transitions at $t'=2t\pm\mu$. We have seen
earlier that at these critical values, the band closes, and simultaneously,
the winding number undergoes a change from $2$ to $1$ at $t'=2t-\mu$,
and from $1$ to $0$ at $t'=2t+\mu$ (Fig.~\ref{fig:4}a). Now, we see
from Fig.~\ref{fig:5} that the entanglement entropy calculated either
via the horizontal or vertical division offers a sharp signature at
each of these phase transitions. For strong inter-leg hopping $t'$ the
two legs are strongly correlated and the system approaches the
maximally entangled state with entanglement entropy $L\ln 2$. One can
think of the system tending to a state consisting of $L$ singlets, one over
each rung of the ladder. When open boundary conditions are imposed
onto the system, degenerate Majorana edge modes also contribute to
entanglement entropy ($S_h$) resulting a starting value of $\ln 2$ rather
than zero for small inter-leg hopping amplitudes ($t'$).

The same behavior can be seen in Fig.~\ref{fig:5}(b) when the system
is divided vertically. In the absence of inter-leg hopping $t'$ the
two legs of the ladder behave as two independent Kitaev chains each
contributing $\ln(2)$ to entanglement entropy.  However, strong
inter-leg hopping $t'$ increases the quantum correlations between the
two legs of the ladder, and weakens the quantum correlations along the
legs of the ladder. This causes the vertical division entanglement
entropy $S_v$ to fall with $t'$. The other striking feature from this
figure is the occurrence of a kink in $S_v$ at the quantum phase
transition. 

\begin{figure}[h!]
\includegraphics[scale=0.7]{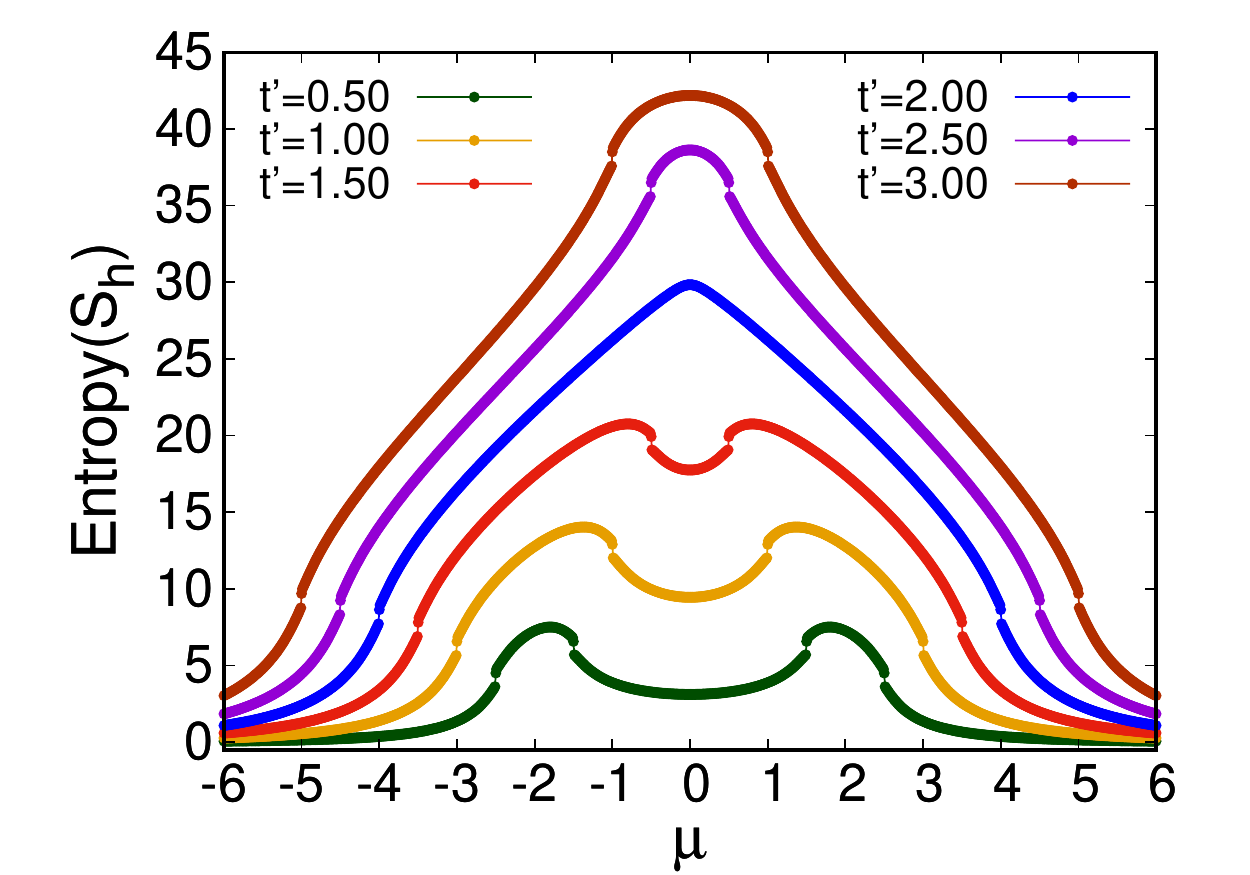}
\caption{The entanglement entropy ($S_h$) in units of $\ln 2$ of Kitaev ladder under horizontal division with various filling $\mu$, for different rungs hopping $t'$. The parameters are set to $L=50$, $t=\Delta=1.0$, $\Phi=0$ and periodic boundary conditions are imposed.}
\label{fig:6}
\end{figure}
In Fig.~\ref{fig:6}, the behavior of the horizontal division
entanglement entropy ($S_h$) with filling $\mu$ is shown. For large
$|\mu|$ the system tends to approach either the fully filled state or
the fully empty state, yielding zero entanglement entropy ($S_h$) in
each case as shown in Fig.~\ref{fig:6}. A discontinuity in the entropy
is seen at the four quantum phase transition points i.e. $|\mu|=2t\pm
t'$. The system possesses two types of quantum phases as described in
Fig.~\ref{fig:4}(a). The entropy features different behavior in
different topological regions. The case $t'=2t$ is special, where two of the discontinuities merge at the origin ($\mu = 0$,
corresponding to half-filling).  We expect a maximum entanglement
entropy close to half filling (i.e. $\mu\sim 0$) for $t'>2t$ as
confirmed in Fig.~\ref{fig:6}.  In contrast, for $t'<2t$ in the
topological region, we see that the entanglement entropy features a
broader flat region close to $\mu~\sim 0$ due to weakened singlets
between the legs of the ladder. This broader region goes to zero for
$t'\to 0$ but under open boundary conditions it starts
from $\ln 2$ due to the presence of degenerate Majorana edge modes.
\begin{figure}[h!]
\includegraphics[scale=0.7]{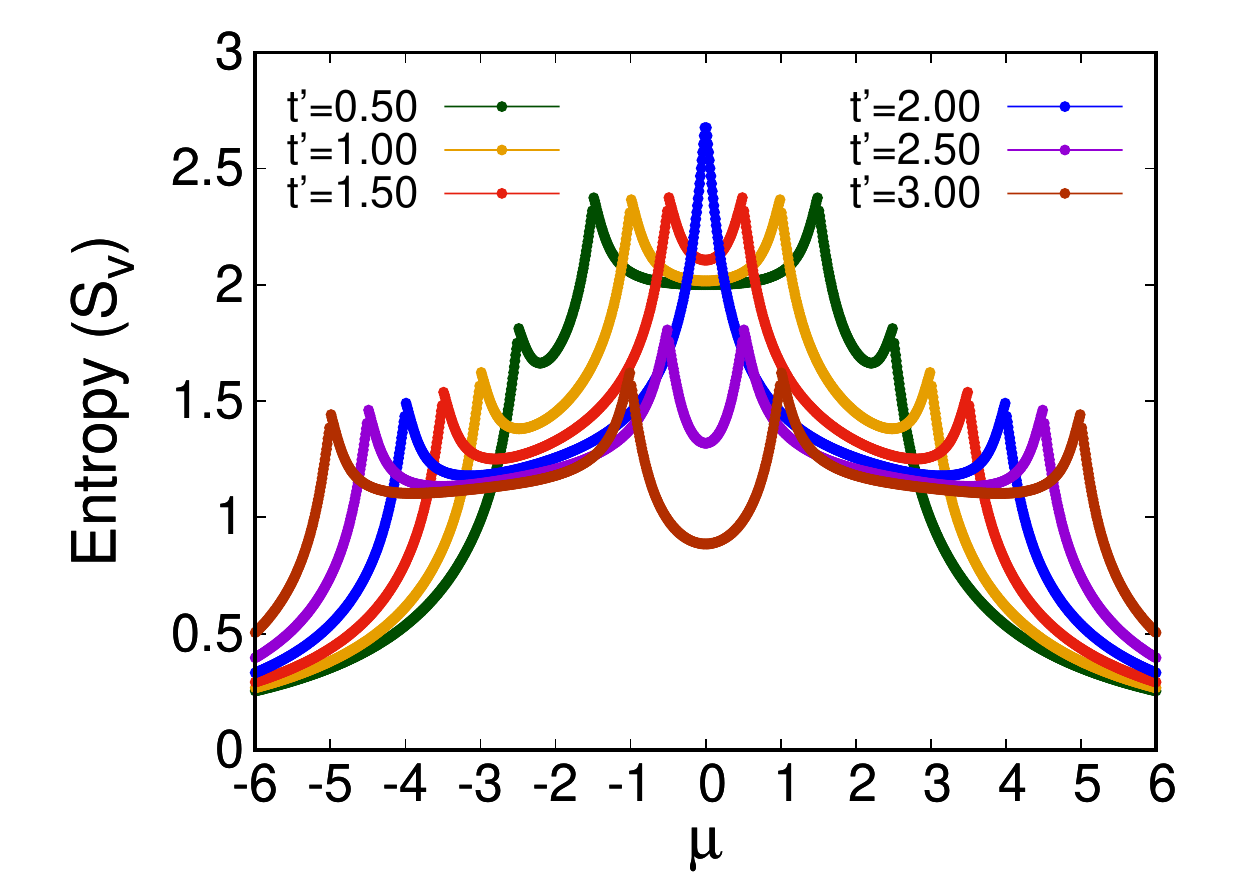}
\caption{The entanglement entropy ($S_v$) in units of $\ln 2$ of Kitaev ladder under vertical division with various filling $\mu$, for different rungs hopping $t'$ for set of parameters $L=50$, $t=\Delta=1.0$, $\phi=0$ and under imposed periodic boundary conditions.}
\label{fig:7}
\end{figure}

The vertical division entanglement entropy ($S_{v}$) between two halves
of the ladder, shown in Fig.~\ref{fig:7}, is also very revealing. The
peaks signal the critical points ($|\mu|=2t\pm t'$) corresponding to
quantum phase transitions. The entanglement entropy ($S_{v}$) close to
half filling ($\mu \sim 0$) signifies the change from $2\ln 2$ to zero when the system
switches from a topological to a trivial phase, just like in the normal Kitaev chain. In
contrast to $S_h$ this entanglement entropy ($S_v$) is small in the limit of
strong inter-leg hopping $t'$ and high in the limit of small $t'$ due to formation of strong quantum correlations
along the legs of the superconducting ladder.
\begin{figure}[h!]
\includegraphics[scale=0.335]{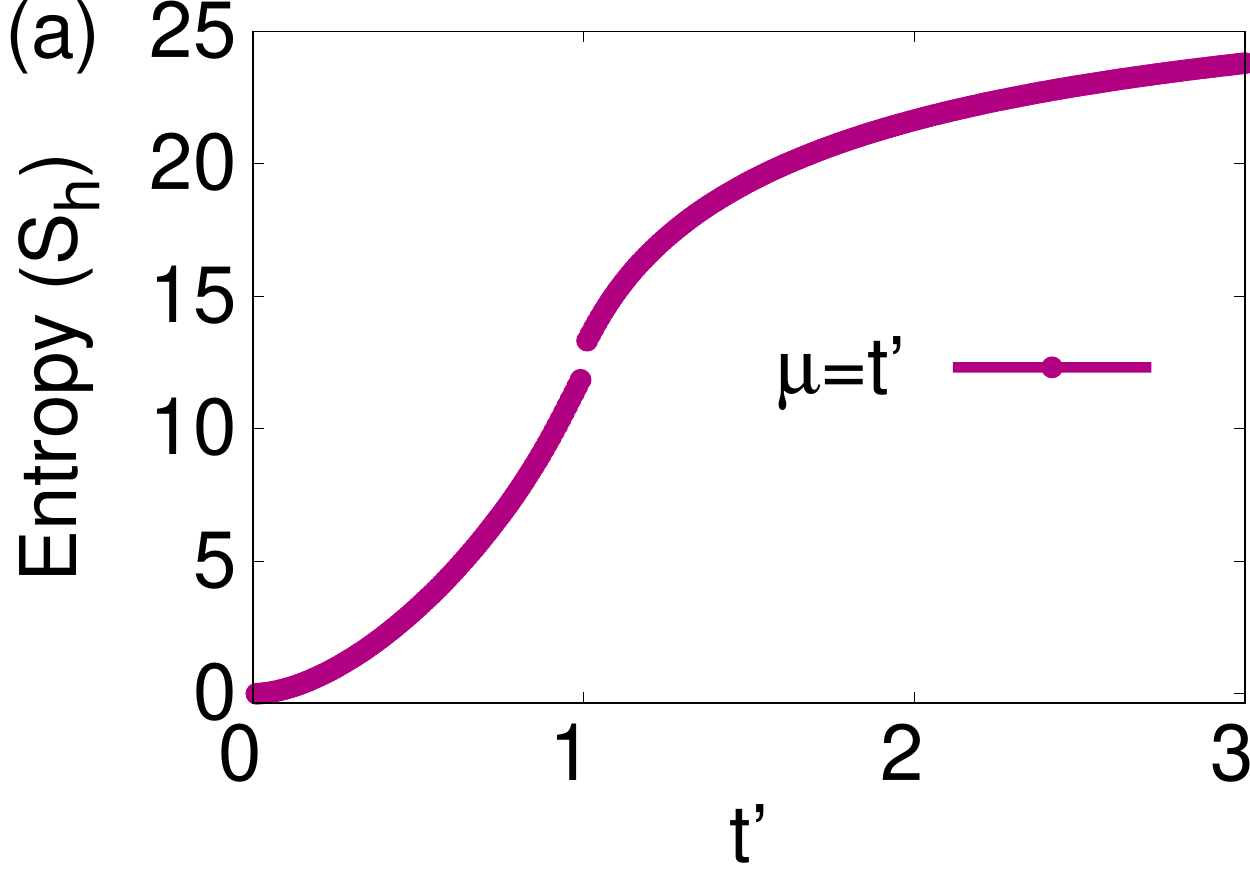}
\includegraphics[scale=0.335]{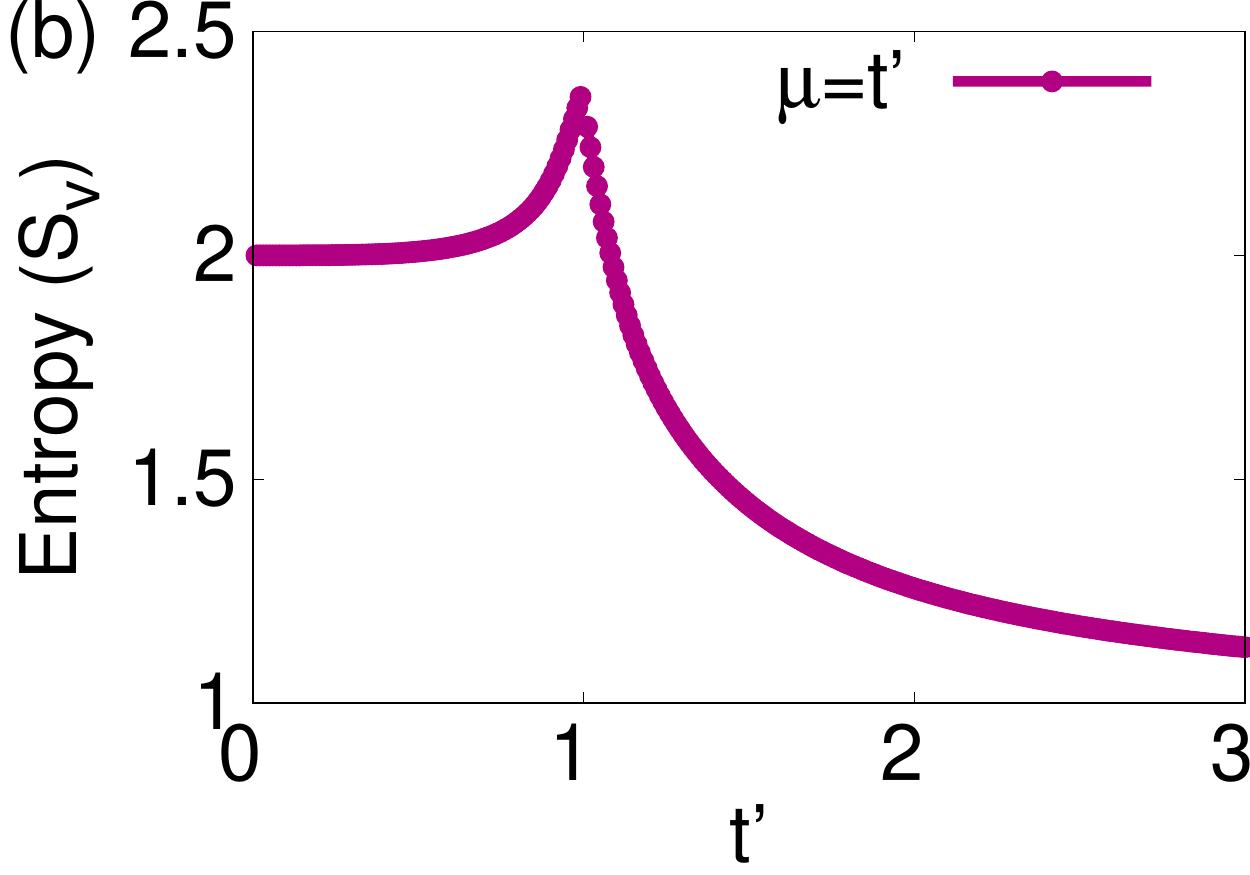}
\caption{The flat band entanglement entropy (a) $S_h$ (b) $S_v$ of the Kitaev ladder with various $t'$ for the set of parameters $L=50$, $t=\Delta=1.0$, $\phi=0$, $t'=\mu$ in units of $\ln 2$.}
\label{fig:8}
\end{figure}

As we know, the flat band is a special case leading to high degeneracy in
the system. In the absence of superconducting phases $\Phi=0$ with
$t=\Delta$ and $\mu=t'$ (in Eq.~\ref{eq14}) the Kitaev ladder possesses two flat bands and two
dispersive bands (Fig.~\ref{dispersion}(a)). The behavior of the entanglement entropy under such
conditions is shown in Fig.~\ref{fig:8} with $\mu=t'$. The flat band
limit is attained by varying filling in the system which results in
low saturation values of $S_h$ as shown in Fig.~\ref{fig:8}(a). There
is only one quantum phase transition point at $t'=t$. The
discontinuity in entanglement entropy in both Fig.~\ref{fig:8}(a,b) is
a clear signatures of this quantum phase transition at $t'=t$.

\subsection{Phase difference \textbf{$\Phi=\pi$}}
\begin{figure}[h!]
\includegraphics[scale=0.35,angle=-90]{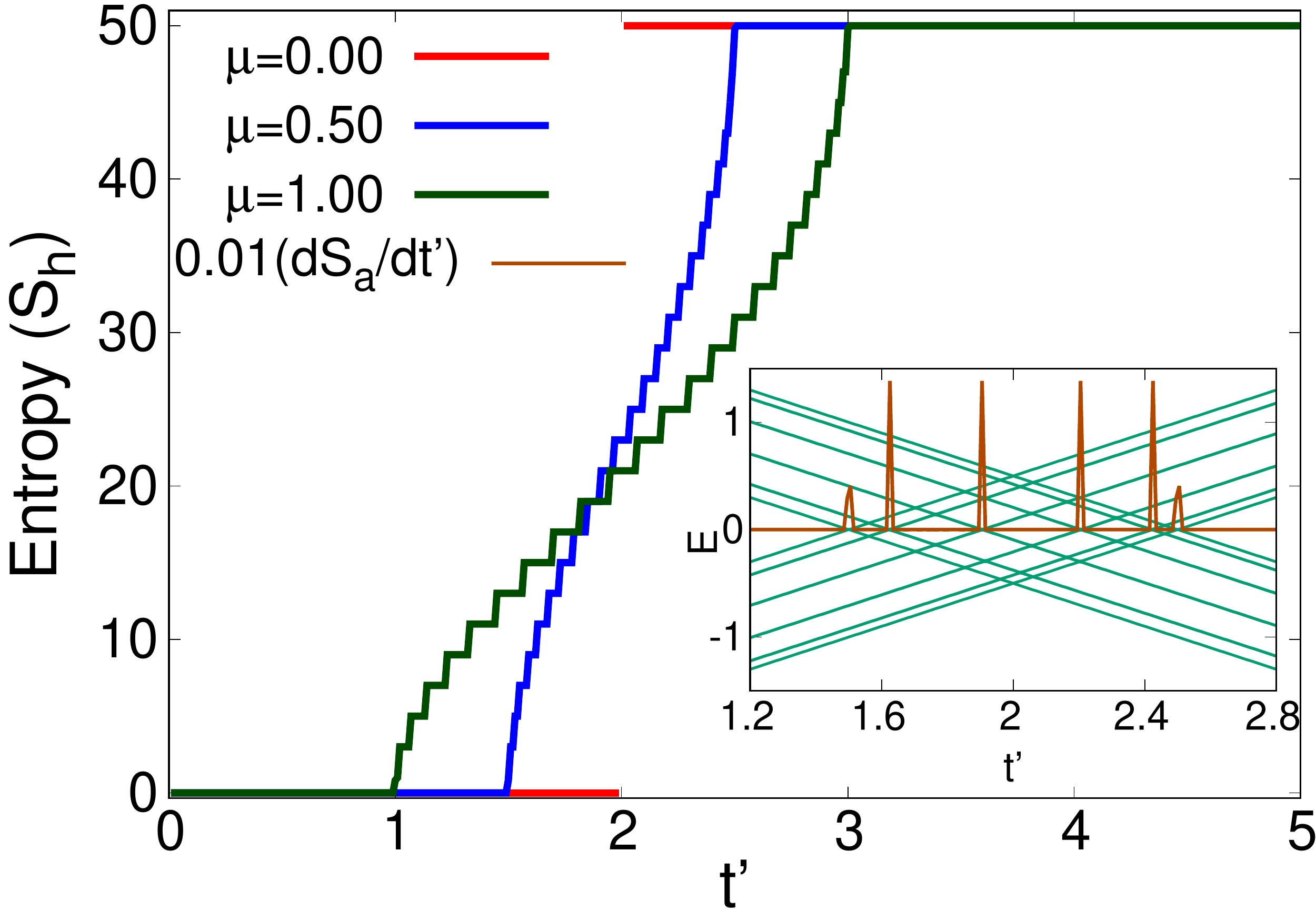}
\caption{The entanglement entropy ($S_h$) of the upper leg of Kitaev ladder with respect to the other leg of the ladder with various $t'$ for  the parameters $L=50$, $t=\Delta=1.0$, $\Phi=\pi\;(\phi_1=0,\phi_2=\pi)$ in units of $\ln 2$. The periodic boundary condtions are imposed to the system. The inset represent variation of energy spectrum and the derivative of entanglement entropy ($S_h$) of Kitaev ladder with the inter-leg hoppong $t'$ for the choice of system parameters $L=10$, $t=\Delta=1$, $\mu=0.5$, $\Phi=\pi$.}
\label{fig:9}
\end{figure}

In this subsection, the superconducting phase difference is maintained
at $\Phi=\pi\;(\phi_1=0,\phi_2=\pi)$. Again the focus of the study will be the two types of
entanglement entropy i.e. $S_h$ and $S_v$. The role of the flat band
that appears on a further fine-tuning of the parameters, is then
studied as a special case. Fig.~\ref{fig:9} illustrates the nature of
the horizontal division entanglement entropy ($S_h$) as a function of the inter-leg
hopping $t'$.

The dispersion relation for the Kitaev ladder with $\Phi=\pi$ is given by
\begin{equation}
\label{eq:pi}E(k)=\pm\Big(\sqrt{(2t\cos k+\mu)^2+4\Delta^2\sin^2k}\pm t'\Big).
\end{equation}
So the Kitaev ladder with $\Phi = \pi$ may be thought of as being
effectively made up of two independent Kitaev chains with the origin
of the energy shifted by $+t'$ in one case and $-t'$ in the other.  We
see from Fig.~\ref{fig:9} that the entanglement entropy remains zero,
until $t'$ hits a threshold value. This feature is to be contrasted
with Fig.~\ref{fig:5}(a) where the entanglement begins to rise for the
tiniest of $t'$. We can understand this feature as a consequence of
the system effectively being two independent chains; entanglement
entropy ($S_h$) remains zero until the energy levels of the two chains
cross each other as can be seen from the inset of Fig.~\ref{fig:9}.

As the coupling strength $t'$ is increased, we see a staircase-like
increase in the entanglement entropy. We are able to identify the
jumps to coincide with a simultaneous crossing of bands in the energy
spectrum as shown in the inset of Fig.~\ref{fig:9}. The point at which
bands cross is also a point of degeneracy.  Each degenerate state
contributes an amount $\ln 2$ to entanglement entropy and so the
magnitude of the jump in $S_h$ is $\ln 2$ times the degeneracy
at the band-crossing point in question. It can be seen that the
derivative~\citep{borchmann2017analytic} of entanglement entropy
nicely captures this feature. The staircase structure starts at
$t'=2t-\mu$ and saturates at $t'=2t+\mu$. Furthermore, the two
critical values are the points where the band closes and the
entanglement entropy nicely captures these quantum phase transitions
of the system. The entanglement entropy saturates at the maximum
possible value of $S_\text{sat}=L\ln 2$ where $L$ is the number of
rungs of the ladder system. We infer that this is due to the formation
of singlets along every rung of the ladder. With open boundary
conditions, degenerate Majorana edge modes are present at $\pm\frac{
  t'}{t}$ separately for each of the two independent Kitaev chains of
the ladder. These Majorana edge modes again yield higher entanglement
entropy ($S_h$) of $\ln 2$ ( as opposed to zero, for periodic boundary
conditions) for small inter-leg hopping amplitudes ($t'$).
\begin{figure}[h!]
\includegraphics[scale=0.7]{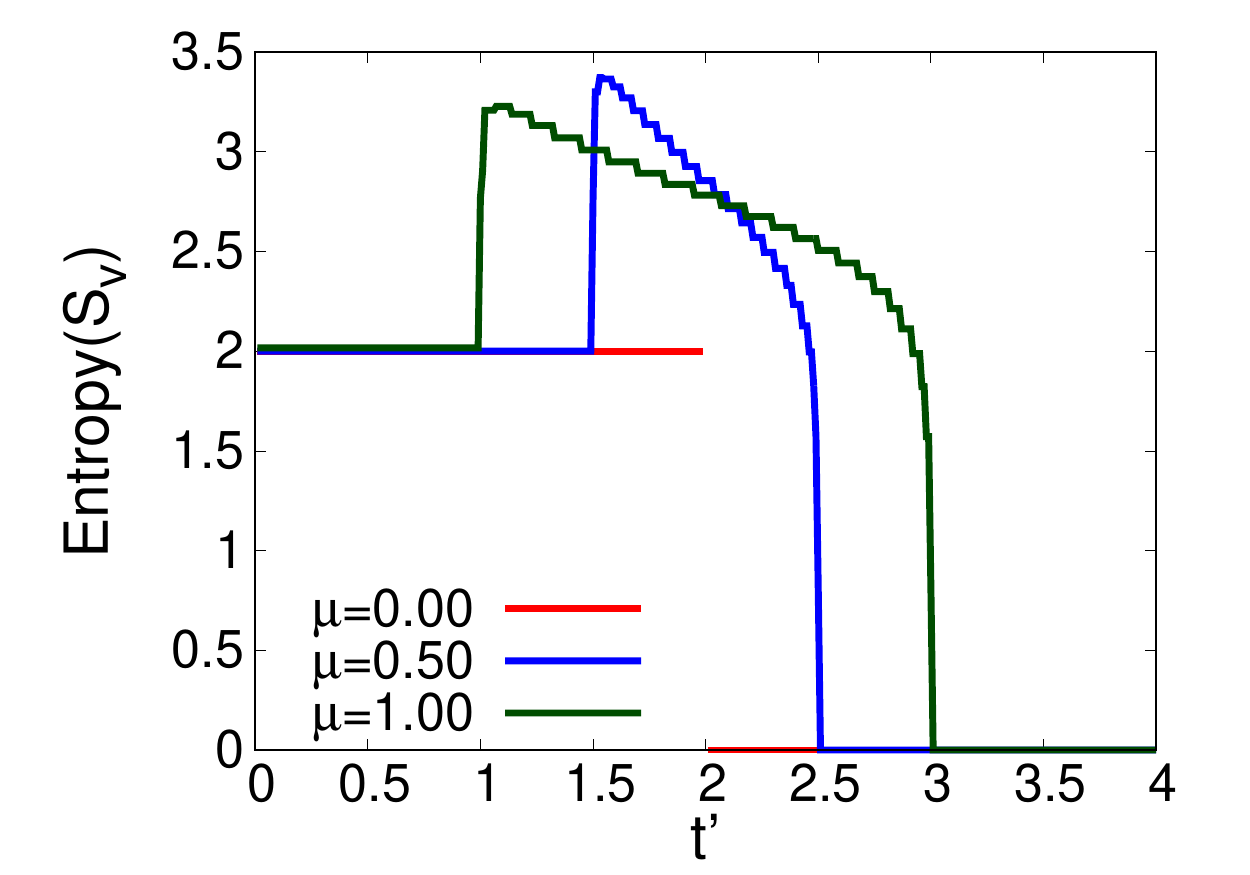}
\caption{The entanglement entropy ($S_v$) in the units of $\ln 2$ of the half Kitaev ladder with respect to the other half of the ladder with various $t'$ for set of parameters $L=50$, $t=\Delta=1.0$, $\Phi=\pi$.}
\label{fig:10}
\end{figure} 

The flat band condition is a special case involving highly degenerate
states in the system. It appears when $t=\Delta$, $\mu = 0$, in
addition to $\Phi=\pi$. For this parameter set, all the four bands are
flat, in contrast to the earlier case where two bands were dispersive
and two were flat (Fig.~\ref{dispersion}). It can be seen from
Fig.~\ref{fig:9} that when the system is tuned to the flat band condition,
the horizontal division entanglement entropy ($S_h$) of the system
shows a large and sudden jump from the minimum value i.e. zero all the
way upto the maximum value i.e. $L\ln 2$ at $t'=2t$. This is due to
the crossing of highly degenerate flat energy bands at $t'=2t$. It is worth noting that for the 
parameter set that we have imposed here, it turns out that each of the
constitutent Kitaev chains itself is subject to its own flat band condition (i.e., $t=\Delta$).

The vertical division entanglement entropy ($S_v$) confirms the
 above observations as shown in Fig.~\ref{fig:10}. Before crossing the
lower critical point $t' = 2t-\mu$, the entanglement entropy ($S_v$) is
locked at a value of $2\ln 2$. Once again, this can be understood if
we think of the Kitaev ladder as being made of two individual Kitaev
chains, each contributing an amount of $\ln 2$ to the entanglement
entropy. Also beyond the higher critical point $t' = 2t+\mu$, the
entanglement entropy becomes zero confirming the formation of a state
made up of singlets along each rung of the ladder.  In between these
critical bounds the entanglement entropy shows discontinuous behavior.
This is once again understood as a consequence of the crossing of bands at 
these points. Once again strikingly, when the system is tuned to the flat band condition,
there is a single big jump in $S_v$ from $2\ln 2$ to zero at the quantum phase transition.

\begin{figure}[h!]
\includegraphics[scale=0.7]{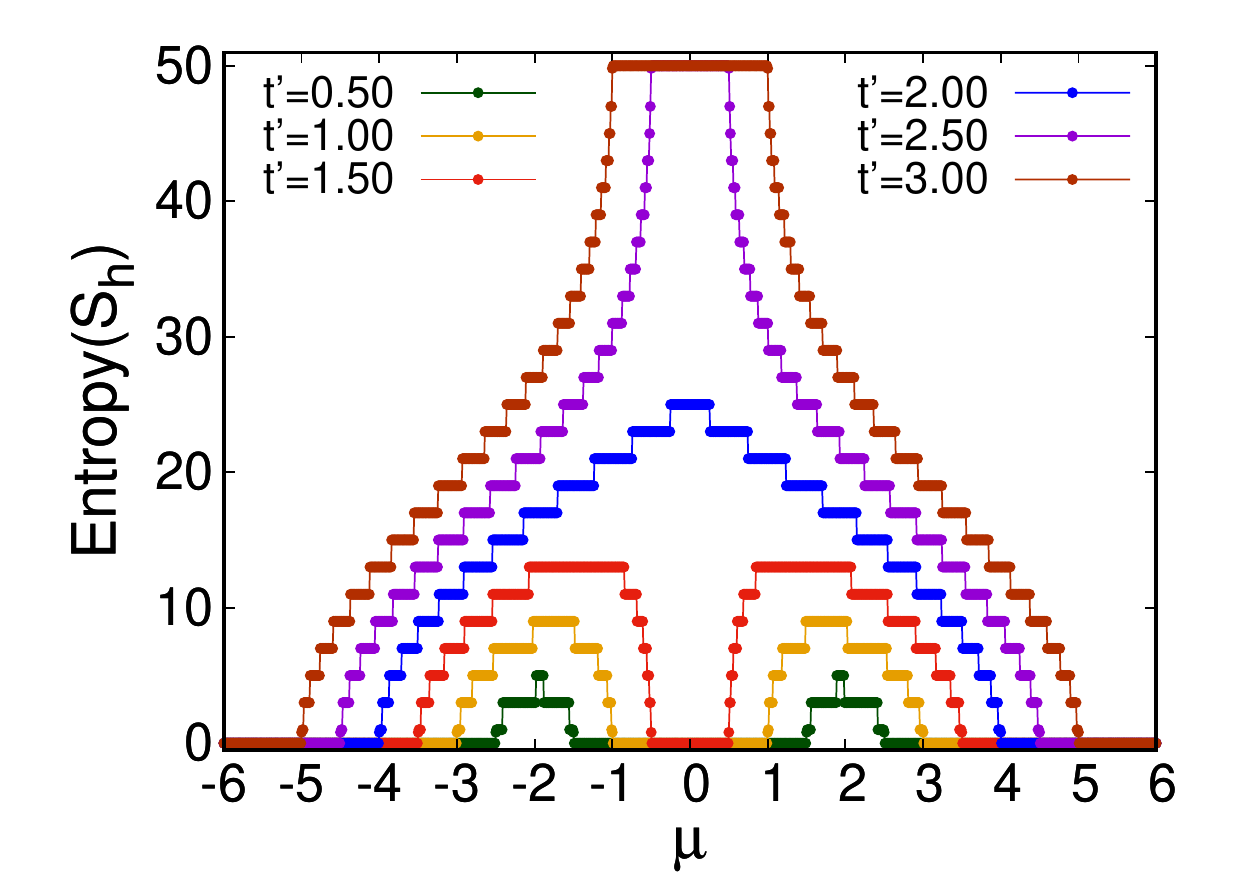}
\caption{The horizontal entanglement entropy ($S_h$) in units of $\ln 2$ of Kitaev ladder with various $\mu$ for parameters $L=50$, $t=\Delta=1.0$, $\Phi=\pi$ and for different $t'$ values. }
\label{fig:11}
\end{figure} 
 
\begin{figure}[h!]
\includegraphics[scale=0.71]{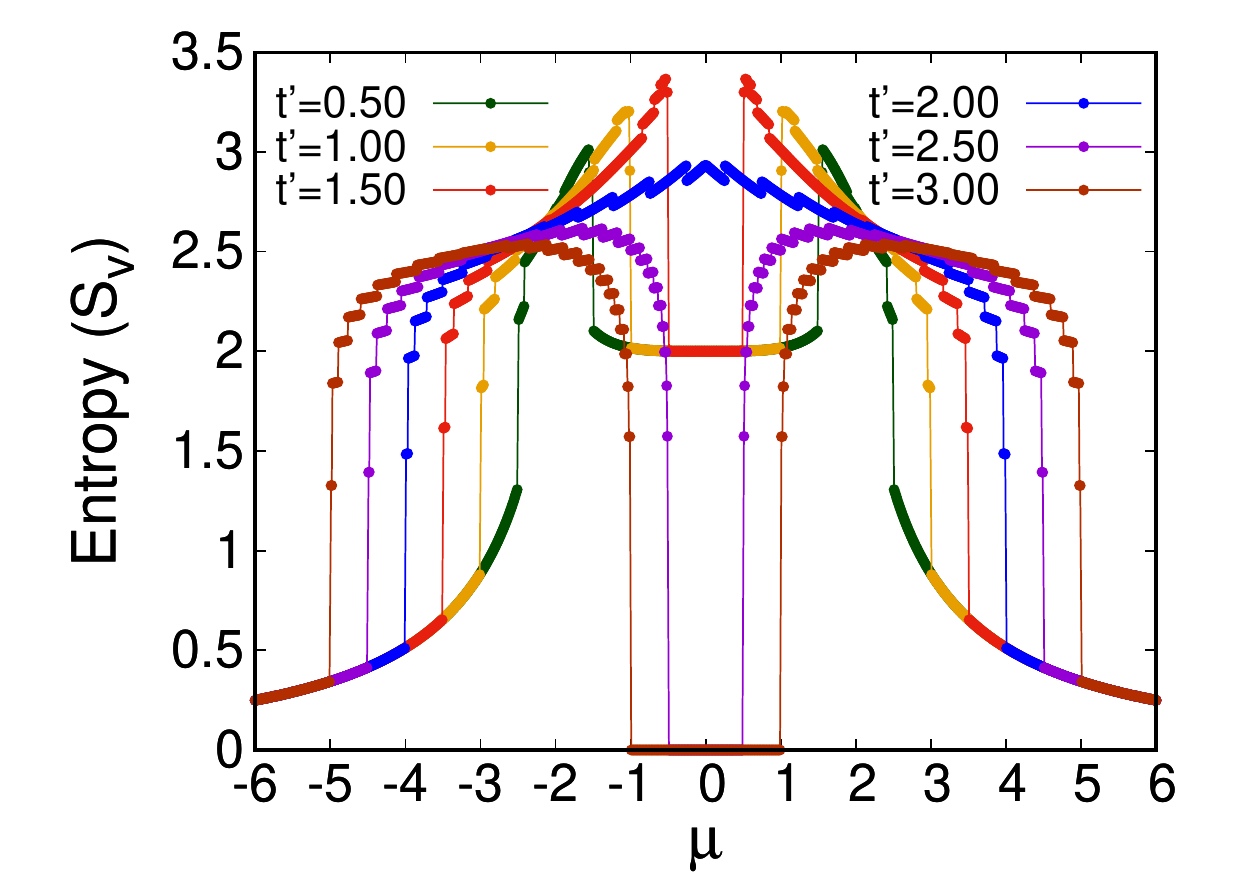}
\caption{The entanglement entropy ($S_v$) of Kitaev ladder under vertical division in units of $\ln 2$ with various filling $\mu$, for different rungs hopping $t'$. The parameters are choosen to be $L=50$, $t=\Delta=1.0$, $\Phi=\pi$ and periodic boundary conditions are imposed. }
\label{fig:12}
\end{figure}

A study of the dependence of the entanglement entropy as a function of
the chemical potential $\mu$ is also instructive. The entanglement
pertaining to horizontal division is shown in Fig.~\ref{fig:11} while
that related to the vertical division is shown in Fig.~\ref{fig:12}.
When the rungs coupling $t'$ is very small, the system behaves as two
independent Kitaev chains and the topological-to-trivial phase
transition occurs at $\mu =\pm(2t\pm t')$ as already shown in Fig.~\ref{fig:4}(b). 

Let us first look at Fig.~\ref{fig:11}. We see that for any value of
$t'$, the entanglement entropy ($S_h$) is a symmetric function of $\mu$
about the origin. We recall from Fig.~\ref{fig:4}(b) that for $\mu=\pm
(2t+t')$, the bands cross, and therefore in the regime $|\mu|>(2t+t')$
where the two chains behave independently, $S_h$ remains zero.  In the
regime $2t-t'<|\mu|<2t+t'$, we see the entanglement grow in a
staircase-like shape, with the jumps happening at points where two
bands cross, and whose jump height is given by $\ln 2$ times the
degeneracy. We observe that $S_h$ first increases and reaches a
maximum value which is characteristic of the particular $t'$ in
question, and has to do with the degree of degeneracy in the
dispersion. It then falls again with the overall curve being symmetric
in $\mu$.  For $\mu=0$, the flat band condition
holds (along with $t=\Delta$ and
$\Phi=\pi$, which have already been imposed in this section), and we see striking behaviour in $S_h$ as the system undergoes
a transition from the topological to the trivial phase. As we have
already seen in Fig.~\ref{fig:9}, $S_h$ undergoes a sudden jump from
zero entropy for $t'<2t$ to $L\ln 2$ entropy for $t'>2t$. In
Fig.~\ref{fig:11}, we see that the size of the flat zero entropy
region around the origin keeps on shrinking as $t'$ approaches $2t$
from below, and then flips over to a flat maximum entropy region
around the orgin, whose size increases with $t'$ away from $2t$.
However, under open boundary conditions close to half filling
($\mu=0$) the entanglement entropy ($S_h$) features a flip from $\ln 2$
to $L\ln 2$ due to the presence of Majorana edge modes.

Next we look at the vertical division entanglement entropy ($S_v$) shown in
Fig.~\ref{fig:12}. Like $S_h$, $S_v$ is also symmetric in $\mu$ about the
origin. In the regime $|\mu|>(2t+t')$ where the two chains behave
independently, $S_v$ is small but not zero, because of weak quantum
correlations along the legs of the ladder. At these conditions $|\mu| =
2t+t'$, and $|\mu| = 2t-t'$, we see jumps in $S_v$, like before. We
observe that in regions where $S_h$ is high, $S_v$ tends to be small,
presumably because of the underlying monogamy ~\citep{YANG2006249} restrictions,
which forbids high entanglement of one subsystem with multiple others.
The special case of $\mu = 0$ corresponding to the flat band condition
displays a sharp dramatic change as the system undergoes a topological
phase transition. In the topological phase $t'<2t$, $S_v = 2\ln 2$,
and is zero beyond $t'>2t$, in the trivial phase. We also observe a
flat zero entropy region close to $\mu = 0$ in the trivial phase, which flips over
to a flat region of value $2\ln 2$ about $\mu = 0$ in the topological phase.

\section{Summary}
In this work, we report the existence of flat bands in a Kitaev ladder
that includes a p-wave superconducting term. These flat bands in
dispersion are associated with strongly localised eigenstates known as
compact localized states (CLS). With the help of a Bogoliubov
transformation, we are able to map the Hamiltonian of the p-wave
superconductor onto an interlinked cross-stitch lattice, and thus
explicitly obtain the underlying compact localized
eigenstates.

Studying the dispersion allows us to propose two different parameter
sets of the Hamiltonian that yield flat bands.  Firstly, we discuss
the presence of flat bands for the Kitaev ladder in the absence of
superconducting phase difference $\Phi=0$ featuring two dispersive and
two flat bands. Next, tuning the superconducting phase difference to
$\Phi=\pi$ results in a band structure with only flat bands leaving
the system completely insulating. The CLS corresponding to these flat
bands are localised on two unit cells of the lattice and with zero
amplitude elsewhere. The high localisation of these states is
responsible for the suppression of transport in the system.

Building on the techniques of Vitagliano et
al~\cite{vitagliano2010volume}, we are able to compute many body
entanglement entropy in the eigenstates of the Kitaev
ladder. Analogous to numerous other quantum phase transitions,
entanglement entropy proves to be a striking diagnostic of the
underlying topological-to-trivial phase transition in this system.
The inclusion of superconducting phase in the Kitaev ladder gives rise
to a change in topological class from BDI to D but entanglement
entropy remains a robust diagnostic in both topological classes. For
the computation of entanglement entropy, we consider two bi-partitions
of the system namely the horizontal and vertical divisions, which
offer complementary insight.

The entanglement entropy is found to strongly depend on the choice of
inter-leg hopping $t'$ in both the partitions of the system. Due to
underlying monogamy constraints, the trends of the horizontal division
entanglement entropy contrast those of the vertical division
entanglement entropy.  In the absence of superconducting phase
difference ($\Phi=0$), strong inter-leg hopping connecting the two
legs of the ladder, tends to enhance the tendency to form singlets
along the rungs, and results in high horizontal division entanglement
entropy. When the parameters are specially tuned (at $t=\Delta$,
$\mu=t'$ and $\Phi=0$) to admit a \emph{flat band}, the horizontal
division entanglement entropy seems to be generally suppressed, in
particular featuring lower maximum values compared to non-flat band
tuning. In the presence of superconducting phase difference
($\Phi=\pi$), the Kitaev ladder can be mapped to
two independent Kitaev chains with shifts in energies by $\frac{\pm
  t'}{t}$. The horizontal division entanglement entropy shows a
staircase behaviour where the sizes of the steps closely match the
degeneracy of the crossing bands. Also, the entanglement entropy
features a sharp jump at the \emph{flat band} condition ($t=\Delta$,
$\mu=0$ and $\Phi=\pi$) due to the crossing of highly degenerate
bands.

We envisage that our work could open up a range of new avenues for
further exploration.  The conventional way to analyse flat bands is to
first discover the compact localised states and then to identify the
corresponding Hamiltonian. Here, we take the opposite direction; we
uncover the existence of flat bands in the very simple Kitaev ladder,
followed by detection of corresponding compact localization states. It
appears that the p-wave pairing symmetry is critical to enabling
flat-bands in the present system. Our own preliminary calculations
(not reported here) indicate that the introduction of the pairing
terms along the rungs of the ladder can also yield flat bands. We
speculate that this method may be general and may help in engineering
other simple geometry lattices that support flat bands. Furthermore,
we believe that the study of many-body entanglement entropy could
offer fresh insights in other systems. For example, the emergence of
Majorana flat bands in s-wave and d-wave
superconductors~\citep{PhysRevB.89.140507,PhysRevB.100.125141} is
connected to their underlying topological properties. A study of
many-body entanglement entropy in such systems may provide a finer
understanding of the topological properties and edge state behaviour
of such lattice structures.

\section{Acknowledgement}
A. S. is grateful to DST for the DST-INSPIRE Faculty Award [DST/INSPIRE/04/2014/002461].
D.S.B acknowledges PhD fellowship support from UGC India.
\bibliography{ref}

\begin{thebibliography}{54}%
\makeatletter
\providecommand \@ifxundefined [1]{%
 \@ifx{#1\undefined}
}%
\providecommand \@ifnum [1]{%
 \ifnum #1\expandafter \@firstoftwo
 \else \expandafter \@secondoftwo
 \fi
}%
\providecommand \@ifx [1]{%
 \ifx #1\expandafter \@firstoftwo
 \else \expandafter \@secondoftwo
 \fi
}%
\providecommand \natexlab [1]{#1}%
\providecommand \enquote  [1]{``#1''}%
\providecommand \bibnamefont  [1]{#1}%
\providecommand \bibfnamefont [1]{#1}%
\providecommand \citenamefont [1]{#1}%
\providecommand \href@noop [0]{\@secondoftwo}%
\providecommand \href [0]{\begingroup \@sanitize@url \@href}%
\providecommand \@href[1]{\@@startlink{#1}\@@href}%
\providecommand \@@href[1]{\endgroup#1\@@endlink}%
\providecommand \@sanitize@url [0]{\catcode `\\12\catcode `\$12\catcode
  `\&12\catcode `\#12\catcode `\^12\catcode `\_12\catcode `\%12\relax}%
\providecommand \@@startlink[1]{}%
\providecommand \@@endlink[0]{}%
\providecommand \url  [0]{\begingroup\@sanitize@url \@url }%
\providecommand \@url [1]{\endgroup\@href {#1}{\urlprefix }}%
\providecommand \urlprefix  [0]{URL }%
\providecommand \Eprint [0]{\href }%
\providecommand \doibase [0]{https://doi.org/}%
\providecommand \selectlanguage [0]{\@gobble}%
\providecommand \bibinfo  [0]{\@secondoftwo}%
\providecommand \bibfield  [0]{\@secondoftwo}%
\providecommand \translation [1]{[#1]}%
\providecommand \BibitemOpen [0]{}%
\providecommand \bibitemStop [0]{}%
\providecommand \bibitemNoStop [0]{.\EOS\space}%
\providecommand \EOS [0]{\spacefactor3000\relax}%
\providecommand \BibitemShut  [1]{\csname bibitem#1\endcsname}%
\let\auto@bib@innerbib\@empty
\bibitem [{\citenamefont {Parameswaran}\ \emph {et~al.}(2013)\citenamefont
  {Parameswaran}, \citenamefont {Roy},\ and\ \citenamefont
  {Sondhi}}]{parameswaran2013fractional}%
  \BibitemOpen
  \bibfield  {author} {\bibinfo {author} {\bibfnamefont {S.~A.}\ \bibnamefont
  {Parameswaran}}, \bibinfo {author} {\bibfnamefont {R.}~\bibnamefont {Roy}},\
  and\ \bibinfo {author} {\bibfnamefont {S.~L.}\ \bibnamefont {Sondhi}},\
  }\bibfield  {title} {\bibinfo {title} {Fractional quantum hall physics in
  topological flat bands},\ }\href@noop {} {\bibfield  {journal} {\bibinfo
  {journal} {Comptes Rendus Physique}\ }\textbf {\bibinfo {volume} {14}},\
  \bibinfo {pages} {816} (\bibinfo {year} {2013})}\BibitemShut {NoStop}%
\bibitem [{\citenamefont {Flach}\ \emph {et~al.}(2014)\citenamefont {Flach},
  \citenamefont {Leykam}, \citenamefont {Bodyfelt}, \citenamefont {Matthies},\
  and\ \citenamefont {Desyatnikov}}]{flach2014detangling}%
  \BibitemOpen
  \bibfield  {author} {\bibinfo {author} {\bibfnamefont {S.}~\bibnamefont
  {Flach}}, \bibinfo {author} {\bibfnamefont {D.}~\bibnamefont {Leykam}},
  \bibinfo {author} {\bibfnamefont {J.~D.}\ \bibnamefont {Bodyfelt}}, \bibinfo
  {author} {\bibfnamefont {P.}~\bibnamefont {Matthies}},\ and\ \bibinfo
  {author} {\bibfnamefont {A.~S.}\ \bibnamefont {Desyatnikov}},\ }\bibfield
  {title} {\bibinfo {title} {Detangling flat bands into fano lattices},\
  }\href@noop {} {\bibfield  {journal} {\bibinfo  {journal} {EPL (Europhysics
  Letters)}\ }\textbf {\bibinfo {volume} {105}},\ \bibinfo {pages} {30001}
  (\bibinfo {year} {2014})}\BibitemShut {NoStop}%
\bibitem [{\citenamefont {Leykam}\ \emph {et~al.}(2018)\citenamefont {Leykam},
  \citenamefont {Andreanov},\ and\ \citenamefont
  {Flach}}]{leykam2018artificial}%
  \BibitemOpen
  \bibfield  {author} {\bibinfo {author} {\bibfnamefont {D.}~\bibnamefont
  {Leykam}}, \bibinfo {author} {\bibfnamefont {A.}~\bibnamefont {Andreanov}},\
  and\ \bibinfo {author} {\bibfnamefont {S.}~\bibnamefont {Flach}},\ }\bibfield
   {title} {\bibinfo {title} {Artificial flat band systems: from lattice models
  to experiments},\ }\href@noop {} {\bibfield  {journal} {\bibinfo  {journal}
  {Advances in Physics: X}\ }\textbf {\bibinfo {volume} {3}},\ \bibinfo {pages}
  {1473052} (\bibinfo {year} {2018})}\BibitemShut {NoStop}%
\bibitem [{\citenamefont {Ashcroft}\ and\ \citenamefont
  {Mermin}(2010)}]{ashcroft2010solid}%
  \BibitemOpen
  \bibfield  {author} {\bibinfo {author} {\bibfnamefont {N.~W.}\ \bibnamefont
  {Ashcroft}}\ and\ \bibinfo {author} {\bibfnamefont {N.~D.}\ \bibnamefont
  {Mermin}},\ }\bibfield  {title} {\bibinfo {title} {Solid state physics
  (saunders college, philadelphia, 1976)},\ }\href@noop {} {\bibfield
  {journal} {\bibinfo  {journal} {Appendix N}\ } (\bibinfo {year}
  {2010})}\BibitemShut {NoStop}%
\bibitem [{\citenamefont {Mielke}(1991)}]{mielke1991ferromagnetism}%
  \BibitemOpen
  \bibfield  {author} {\bibinfo {author} {\bibfnamefont {A.}~\bibnamefont
  {Mielke}},\ }\bibfield  {title} {\bibinfo {title} {Ferromagnetism in the
  hubbard model on line graphs and further considerations},\ }\href@noop {}
  {\bibfield  {journal} {\bibinfo  {journal} {Journal of Physics A:
  Mathematical and General}\ }\textbf {\bibinfo {volume} {24}},\ \bibinfo
  {pages} {3311} (\bibinfo {year} {1991})}\BibitemShut {NoStop}%
\bibitem [{\citenamefont {Tasaki}(1992)}]{tasaki1992ferromagnetism}%
  \BibitemOpen
  \bibfield  {author} {\bibinfo {author} {\bibfnamefont {H.}~\bibnamefont
  {Tasaki}},\ }\bibfield  {title} {\bibinfo {title} {Ferromagnetism in the
  hubbard models with degenerate single-electron ground states},\ }\href@noop
  {} {\bibfield  {journal} {\bibinfo  {journal} {Physical review letters}\
  }\textbf {\bibinfo {volume} {69}},\ \bibinfo {pages} {1608} (\bibinfo {year}
  {1992})}\BibitemShut {NoStop}%
\bibitem [{\citenamefont {Mielke}(1999)}]{mielke1999ferromagnetism}%
  \BibitemOpen
  \bibfield  {author} {\bibinfo {author} {\bibfnamefont {A.}~\bibnamefont
  {Mielke}},\ }\bibfield  {title} {\bibinfo {title} {Ferromagnetism in
  single-band hubbard models with a partially flat band},\ }\href@noop {}
  {\bibfield  {journal} {\bibinfo  {journal} {Physical review letters}\
  }\textbf {\bibinfo {volume} {82}},\ \bibinfo {pages} {4312} (\bibinfo {year}
  {1999})}\BibitemShut {NoStop}%
\bibitem [{\citenamefont {Tasaki}(1998)}]{tasaki1998nagaoka}%
  \BibitemOpen
  \bibfield  {author} {\bibinfo {author} {\bibfnamefont {H.}~\bibnamefont
  {Tasaki}},\ }\bibfield  {title} {\bibinfo {title} {From nagaoka's
  ferromagnetism to flat-band ferromagnetism and beyond: An introduction to
  ferromagnetism in the hubbard model},\ }\href@noop {} {\bibfield  {journal}
  {\bibinfo  {journal} {Progress of Theoretical Physics}\ }\textbf {\bibinfo
  {volume} {99}},\ \bibinfo {pages} {489} (\bibinfo {year} {1998})}\BibitemShut
  {NoStop}%
\bibitem [{\citenamefont {Guzm{\'a}n-Silva}\ \emph {et~al.}(2014)\citenamefont
  {Guzm{\'a}n-Silva}, \citenamefont {Mej{\'\i}a-Cort{\'e}s}, \citenamefont
  {Bandres}, \citenamefont {Rechtsman}, \citenamefont {Weimann}, \citenamefont
  {Nolte}, \citenamefont {Segev}, \citenamefont {Szameit},\ and\ \citenamefont
  {Vicencio}}]{guzman2014experimental}%
  \BibitemOpen
  \bibfield  {author} {\bibinfo {author} {\bibfnamefont {D.}~\bibnamefont
  {Guzm{\'a}n-Silva}}, \bibinfo {author} {\bibfnamefont {C.}~\bibnamefont
  {Mej{\'\i}a-Cort{\'e}s}}, \bibinfo {author} {\bibfnamefont {M.}~\bibnamefont
  {Bandres}}, \bibinfo {author} {\bibfnamefont {M.~C.}\ \bibnamefont
  {Rechtsman}}, \bibinfo {author} {\bibfnamefont {S.}~\bibnamefont {Weimann}},
  \bibinfo {author} {\bibfnamefont {S.}~\bibnamefont {Nolte}}, \bibinfo
  {author} {\bibfnamefont {M.}~\bibnamefont {Segev}}, \bibinfo {author}
  {\bibfnamefont {A.}~\bibnamefont {Szameit}},\ and\ \bibinfo {author}
  {\bibfnamefont {R.}~\bibnamefont {Vicencio}},\ }\bibfield  {title} {\bibinfo
  {title} {Experimental observation of bulk and edge transport in photonic lieb
  lattices},\ }\href@noop {} {\bibfield  {journal} {\bibinfo  {journal} {New
  Journal of Physics}\ }\textbf {\bibinfo {volume} {16}},\ \bibinfo {pages}
  {063061} (\bibinfo {year} {2014})}\BibitemShut {NoStop}%
\bibitem [{\citenamefont {Vicencio}\ \emph {et~al.}(2015)\citenamefont
  {Vicencio}, \citenamefont {Cantillano}, \citenamefont {Morales-Inostroza},
  \citenamefont {Real}, \citenamefont {Mej{\'\i}a-Cort{\'e}s}, \citenamefont
  {Weimann}, \citenamefont {Szameit},\ and\ \citenamefont
  {Molina}}]{vicencio2015observation}%
  \BibitemOpen
  \bibfield  {author} {\bibinfo {author} {\bibfnamefont {R.~A.}\ \bibnamefont
  {Vicencio}}, \bibinfo {author} {\bibfnamefont {C.}~\bibnamefont
  {Cantillano}}, \bibinfo {author} {\bibfnamefont {L.}~\bibnamefont
  {Morales-Inostroza}}, \bibinfo {author} {\bibfnamefont {B.}~\bibnamefont
  {Real}}, \bibinfo {author} {\bibfnamefont {C.}~\bibnamefont
  {Mej{\'\i}a-Cort{\'e}s}}, \bibinfo {author} {\bibfnamefont {S.}~\bibnamefont
  {Weimann}}, \bibinfo {author} {\bibfnamefont {A.}~\bibnamefont {Szameit}},\
  and\ \bibinfo {author} {\bibfnamefont {M.~I.}\ \bibnamefont {Molina}},\
  }\bibfield  {title} {\bibinfo {title} {Observation of localized states in
  lieb photonic lattices},\ }\href@noop {} {\bibfield  {journal} {\bibinfo
  {journal} {Physical review letters}\ }\textbf {\bibinfo {volume} {114}},\
  \bibinfo {pages} {245503} (\bibinfo {year} {2015})}\BibitemShut {NoStop}%
\bibitem [{\citenamefont {Mukherjee}\ \emph {et~al.}(2015)\citenamefont
  {Mukherjee}, \citenamefont {Spracklen}, \citenamefont {Choudhury},
  \citenamefont {Goldman}, \citenamefont {{\"O}hberg}, \citenamefont
  {Andersson},\ and\ \citenamefont {Thomson}}]{mukherjee2015observation}%
  \BibitemOpen
  \bibfield  {author} {\bibinfo {author} {\bibfnamefont {S.}~\bibnamefont
  {Mukherjee}}, \bibinfo {author} {\bibfnamefont {A.}~\bibnamefont
  {Spracklen}}, \bibinfo {author} {\bibfnamefont {D.}~\bibnamefont
  {Choudhury}}, \bibinfo {author} {\bibfnamefont {N.}~\bibnamefont {Goldman}},
  \bibinfo {author} {\bibfnamefont {P.}~\bibnamefont {{\"O}hberg}}, \bibinfo
  {author} {\bibfnamefont {E.}~\bibnamefont {Andersson}},\ and\ \bibinfo
  {author} {\bibfnamefont {R.~R.}\ \bibnamefont {Thomson}},\ }\bibfield
  {title} {\bibinfo {title} {Observation of a localized flat-band state in a
  photonic lieb lattice},\ }\href@noop {} {\bibfield  {journal} {\bibinfo
  {journal} {Physical review letters}\ }\textbf {\bibinfo {volume} {114}},\
  \bibinfo {pages} {245504} (\bibinfo {year} {2015})}\BibitemShut {NoStop}%
\bibitem [{\citenamefont {Zhang}\ and\ \citenamefont
  {Zhang}(2013)}]{zhang2013bose}%
  \BibitemOpen
  \bibfield  {author} {\bibinfo {author} {\bibfnamefont {Y.}~\bibnamefont
  {Zhang}}\ and\ \bibinfo {author} {\bibfnamefont {C.}~\bibnamefont {Zhang}},\
  }\bibfield  {title} {\bibinfo {title} {Bose-einstein condensates in
  spin-orbit-coupled optical lattices: Flat bands and superfluidity},\
  }\href@noop {} {\bibfield  {journal} {\bibinfo  {journal} {Physical Review
  A}\ }\textbf {\bibinfo {volume} {87}},\ \bibinfo {pages} {023611} (\bibinfo
  {year} {2013})}\BibitemShut {NoStop}%
\bibitem [{\citenamefont {Gladchenko}\ \emph {et~al.}(2009)\citenamefont
  {Gladchenko}, \citenamefont {Olaya}, \citenamefont {Dupont-Ferrier},
  \citenamefont {Dou{\c{c}}ot}, \citenamefont {Ioffe},\ and\ \citenamefont
  {Gershenson}}]{gladchenko2009superconducting}%
  \BibitemOpen
  \bibfield  {author} {\bibinfo {author} {\bibfnamefont {S.}~\bibnamefont
  {Gladchenko}}, \bibinfo {author} {\bibfnamefont {D.}~\bibnamefont {Olaya}},
  \bibinfo {author} {\bibfnamefont {E.}~\bibnamefont {Dupont-Ferrier}},
  \bibinfo {author} {\bibfnamefont {B.}~\bibnamefont {Dou{\c{c}}ot}}, \bibinfo
  {author} {\bibfnamefont {L.~B.}\ \bibnamefont {Ioffe}},\ and\ \bibinfo
  {author} {\bibfnamefont {M.~E.}\ \bibnamefont {Gershenson}},\ }\bibfield
  {title} {\bibinfo {title} {Superconducting nanocircuits for topologically
  protected qubits},\ }\href@noop {} {\bibfield  {journal} {\bibinfo  {journal}
  {Nature Physics}\ }\textbf {\bibinfo {volume} {5}},\ \bibinfo {pages} {48}
  (\bibinfo {year} {2009})}\BibitemShut {NoStop}%
\bibitem [{\citenamefont {Maimaiti}\ \emph {et~al.}(2017)\citenamefont
  {Maimaiti}, \citenamefont {Andreanov}, \citenamefont {Park}, \citenamefont
  {Gendelman},\ and\ \citenamefont {Flach}}]{maimaiti2017compact}%
  \BibitemOpen
  \bibfield  {author} {\bibinfo {author} {\bibfnamefont {W.}~\bibnamefont
  {Maimaiti}}, \bibinfo {author} {\bibfnamefont {A.}~\bibnamefont {Andreanov}},
  \bibinfo {author} {\bibfnamefont {H.~C.}\ \bibnamefont {Park}}, \bibinfo
  {author} {\bibfnamefont {O.}~\bibnamefont {Gendelman}},\ and\ \bibinfo
  {author} {\bibfnamefont {S.}~\bibnamefont {Flach}},\ }\bibfield  {title}
  {\bibinfo {title} {Compact localized states and flat-band generators in one
  dimension},\ }\href@noop {} {\bibfield  {journal} {\bibinfo  {journal}
  {Physical Review B}\ }\textbf {\bibinfo {volume} {95}},\ \bibinfo {pages}
  {115135} (\bibinfo {year} {2017})}\BibitemShut {NoStop}%
\bibitem [{\citenamefont {Ramachandran}\ \emph {et~al.}(2017)\citenamefont
  {Ramachandran}, \citenamefont {Andreanov},\ and\ \citenamefont
  {Flach}}]{ramachandran2017chiral}%
  \BibitemOpen
  \bibfield  {author} {\bibinfo {author} {\bibfnamefont {A.}~\bibnamefont
  {Ramachandran}}, \bibinfo {author} {\bibfnamefont {A.}~\bibnamefont
  {Andreanov}},\ and\ \bibinfo {author} {\bibfnamefont {S.}~\bibnamefont
  {Flach}},\ }\bibfield  {title} {\bibinfo {title} {Chiral flat bands:
  Existence, engineering, and stability},\ }\href@noop {} {\bibfield  {journal}
  {\bibinfo  {journal} {Physical Review B}\ }\textbf {\bibinfo {volume} {96}},\
  \bibinfo {pages} {161104} (\bibinfo {year} {2017})}\BibitemShut {NoStop}%
\bibitem [{\citenamefont {Morales-Inostroza}\ and\ \citenamefont
  {Vicencio}(2016)}]{morales2016simple}%
  \BibitemOpen
  \bibfield  {author} {\bibinfo {author} {\bibfnamefont {L.}~\bibnamefont
  {Morales-Inostroza}}\ and\ \bibinfo {author} {\bibfnamefont {R.~A.}\
  \bibnamefont {Vicencio}},\ }\bibfield  {title} {\bibinfo {title} {Simple
  method to construct flat-band lattices},\ }\href@noop {} {\bibfield
  {journal} {\bibinfo  {journal} {Physical Review A}\ }\textbf {\bibinfo
  {volume} {94}},\ \bibinfo {pages} {043831} (\bibinfo {year}
  {2016})}\BibitemShut {NoStop}%
\bibitem [{\citenamefont {Maimaiti}\ \emph {et~al.}(2019)\citenamefont
  {Maimaiti}, \citenamefont {Flach},\ and\ \citenamefont
  {Andreanov}}]{maimaiti2019universal}%
  \BibitemOpen
  \bibfield  {author} {\bibinfo {author} {\bibfnamefont {W.}~\bibnamefont
  {Maimaiti}}, \bibinfo {author} {\bibfnamefont {S.}~\bibnamefont {Flach}},\
  and\ \bibinfo {author} {\bibfnamefont {A.}~\bibnamefont {Andreanov}},\
  }\bibfield  {title} {\bibinfo {title} {Universal d= 1 flat band generator
  from compact localized states},\ }\href@noop {} {\bibfield  {journal}
  {\bibinfo  {journal} {Physical Review B}\ }\textbf {\bibinfo {volume} {99}},\
  \bibinfo {pages} {125129} (\bibinfo {year} {2019})}\BibitemShut {NoStop}%
\bibitem [{\citenamefont {Maiellaro}\ \emph {et~al.}(2018)\citenamefont
  {Maiellaro}, \citenamefont {Romeo},\ and\ \citenamefont
  {Citro}}]{Maiellaro2018}%
  \BibitemOpen
  \bibfield  {author} {\bibinfo {author} {\bibfnamefont {A.}~\bibnamefont
  {Maiellaro}}, \bibinfo {author} {\bibfnamefont {F.}~\bibnamefont {Romeo}},\
  and\ \bibinfo {author} {\bibfnamefont {R.}~\bibnamefont {Citro}},\ }\bibfield
   {title} {\bibinfo {title} {Topological phase diagram of a kitaev ladder},\
  }\href {https://doi.org/10.1140/epjst/e2018-800090-y} {\bibfield  {journal}
  {\bibinfo  {journal} {The European Physical Journal Special Topics}\ }\textbf
  {\bibinfo {volume} {227}},\ \bibinfo {pages} {1397} (\bibinfo {year}
  {2018})}\BibitemShut {NoStop}%
\bibitem [{\citenamefont {Kitaev}(2001)}]{kitaev2001unpaired}%
  \BibitemOpen
  \bibfield  {author} {\bibinfo {author} {\bibfnamefont {A.~Y.}\ \bibnamefont
  {Kitaev}},\ }\bibfield  {title} {\bibinfo {title} {Unpaired majorana fermions
  in quantum wires},\ }\href {https://doi.org/10.1070/1063-7869/44/10s/s29}
  {\bibfield  {journal} {\bibinfo  {journal} {Phys.-Usp.}\ }\textbf {\bibinfo
  {volume} {44}},\ \bibinfo {pages} {131} (\bibinfo {year} {2001})}\BibitemShut
  {NoStop}%
\bibitem [{\citenamefont {Wu}(2012{\natexlab{a}})}]{wu2012topological}%
  \BibitemOpen
  \bibfield  {author} {\bibinfo {author} {\bibfnamefont {N.}~\bibnamefont
  {Wu}},\ }\bibfield  {title} {\bibinfo {title} {Topological phases of the
  two-leg kitaev ladder},\ }\href@noop {} {\bibfield  {journal} {\bibinfo
  {journal} {Physics Letters A}\ }\textbf {\bibinfo {volume} {376}},\ \bibinfo
  {pages} {3530} (\bibinfo {year} {2012}{\natexlab{a}})}\BibitemShut {NoStop}%
\bibitem [{\citenamefont {Nehra}\ \emph
  {et~al.}(2019{\natexlab{a}})\citenamefont {Nehra}, \citenamefont {Bhakuni},
  \citenamefont {Sharma},\ and\ \citenamefont {Soori}}]{nehra2019enhancement}%
  \BibitemOpen
  \bibfield  {author} {\bibinfo {author} {\bibfnamefont {R.}~\bibnamefont
  {Nehra}}, \bibinfo {author} {\bibfnamefont {D.~S.}\ \bibnamefont {Bhakuni}},
  \bibinfo {author} {\bibfnamefont {A.}~\bibnamefont {Sharma}},\ and\ \bibinfo
  {author} {\bibfnamefont {A.}~\bibnamefont {Soori}},\ }\bibfield  {title}
  {\bibinfo {title} {Enhancement of crossed andreev reflection in a kitaev
  ladder connected to normal metal leads},\ }\href@noop {} {\bibfield
  {journal} {\bibinfo  {journal} {Journal of Physics: Condensed Matter}\
  }\textbf {\bibinfo {volume} {31}},\ \bibinfo {pages} {345304} (\bibinfo
  {year} {2019}{\natexlab{a}})}\BibitemShut {NoStop}%
\bibitem [{\citenamefont {Flammia}\ \emph {et~al.}(2009)\citenamefont
  {Flammia}, \citenamefont {Hamma}, \citenamefont {Hughes},\ and\ \citenamefont
  {Wen}}]{flammia2009topological}%
  \BibitemOpen
  \bibfield  {author} {\bibinfo {author} {\bibfnamefont {S.~T.}\ \bibnamefont
  {Flammia}}, \bibinfo {author} {\bibfnamefont {A.}~\bibnamefont {Hamma}},
  \bibinfo {author} {\bibfnamefont {T.~L.}\ \bibnamefont {Hughes}},\ and\
  \bibinfo {author} {\bibfnamefont {X.-G.}\ \bibnamefont {Wen}},\ }\bibfield
  {title} {\bibinfo {title} {Topological entanglement r{\'e}nyi entropy and
  reduced density matrix structure},\ }\href@noop {} {\bibfield  {journal}
  {\bibinfo  {journal} {Physical review letters}\ }\textbf {\bibinfo {volume}
  {103}},\ \bibinfo {pages} {261601} (\bibinfo {year} {2009})}\BibitemShut
  {NoStop}%
\bibitem [{\citenamefont {Nehra}\ \emph
  {et~al.}(2018{\natexlab{a}})\citenamefont {Nehra}, \citenamefont {Bhakuni},
  \citenamefont {Gangadharaiah},\ and\ \citenamefont
  {Sharma}}]{PhysRevB.98.045120}%
  \BibitemOpen
  \bibfield  {author} {\bibinfo {author} {\bibfnamefont {R.}~\bibnamefont
  {Nehra}}, \bibinfo {author} {\bibfnamefont {D.~S.}\ \bibnamefont {Bhakuni}},
  \bibinfo {author} {\bibfnamefont {S.}~\bibnamefont {Gangadharaiah}},\ and\
  \bibinfo {author} {\bibfnamefont {A.}~\bibnamefont {Sharma}},\ }\bibfield
  {title} {\bibinfo {title} {Many-body entanglement in a topological chiral
  ladder},\ }\href {https://doi.org/10.1103/PhysRevB.98.045120} {\bibfield
  {journal} {\bibinfo  {journal} {Phys. Rev. B}\ }\textbf {\bibinfo {volume}
  {98}},\ \bibinfo {pages} {045120} (\bibinfo {year}
  {2018}{\natexlab{a}})}\BibitemShut {NoStop}%
\bibitem [{\citenamefont {Chen}\ and\ \citenamefont
  {Fradkin}(2013)}]{chen2013quantum}%
  \BibitemOpen
  \bibfield  {author} {\bibinfo {author} {\bibfnamefont {X.}~\bibnamefont
  {Chen}}\ and\ \bibinfo {author} {\bibfnamefont {E.}~\bibnamefont {Fradkin}},\
  }\bibfield  {title} {\bibinfo {title} {Quantum entanglement and thermal
  reduced density matrices in fermion and spin systems on ladders},\
  }\href@noop {} {\bibfield  {journal} {\bibinfo  {journal} {Journal of
  Statistical Mechanics: Theory and Experiment}\ }\textbf {\bibinfo {volume}
  {2013}},\ \bibinfo {pages} {P08013} (\bibinfo {year} {2013})}\BibitemShut
  {NoStop}%
\bibitem [{\citenamefont {Eisert}\ \emph {et~al.}(2010)\citenamefont {Eisert},
  \citenamefont {Cramer},\ and\ \citenamefont {Plenio}}]{RevModPhys.82.277}%
  \BibitemOpen
  \bibfield  {author} {\bibinfo {author} {\bibfnamefont {J.}~\bibnamefont
  {Eisert}}, \bibinfo {author} {\bibfnamefont {M.}~\bibnamefont {Cramer}},\
  and\ \bibinfo {author} {\bibfnamefont {M.~B.}\ \bibnamefont {Plenio}},\
  }\bibfield  {title} {\bibinfo {title} {Colloquium: Area laws for the
  entanglement entropy},\ }\href {https://doi.org/10.1103/RevModPhys.82.277}
  {\bibfield  {journal} {\bibinfo  {journal} {Rev. Mod. Phys.}\ }\textbf
  {\bibinfo {volume} {82}},\ \bibinfo {pages} {277} (\bibinfo {year}
  {2010})}\BibitemShut {NoStop}%
\bibitem [{\citenamefont {Kitaev}\ and\ \citenamefont
  {Preskill}(2006)}]{kitaev2006topological}%
  \BibitemOpen
  \bibfield  {author} {\bibinfo {author} {\bibfnamefont {A.}~\bibnamefont
  {Kitaev}}\ and\ \bibinfo {author} {\bibfnamefont {J.}~\bibnamefont
  {Preskill}},\ }\bibfield  {title} {\bibinfo {title} {Topological entanglement
  entropy},\ }\href@noop {} {\bibfield  {journal} {\bibinfo  {journal}
  {Physical review letters}\ }\textbf {\bibinfo {volume} {96}},\ \bibinfo
  {pages} {110404} (\bibinfo {year} {2006})}\BibitemShut {NoStop}%
\bibitem [{\citenamefont {Vodola}\ \emph {et~al.}(2014)\citenamefont {Vodola},
  \citenamefont {Lepori}, \citenamefont {Ercolessi}, \citenamefont {Gorshkov},\
  and\ \citenamefont {Pupillo}}]{PhysRevLett.113.156402}%
  \BibitemOpen
  \bibfield  {author} {\bibinfo {author} {\bibfnamefont {D.}~\bibnamefont
  {Vodola}}, \bibinfo {author} {\bibfnamefont {L.}~\bibnamefont {Lepori}},
  \bibinfo {author} {\bibfnamefont {E.}~\bibnamefont {Ercolessi}}, \bibinfo
  {author} {\bibfnamefont {A.~V.}\ \bibnamefont {Gorshkov}},\ and\ \bibinfo
  {author} {\bibfnamefont {G.}~\bibnamefont {Pupillo}},\ }\bibfield  {title}
  {\bibinfo {title} {Kitaev chains with long-range pairing},\ }\href
  {https://doi.org/10.1103/PhysRevLett.113.156402} {\bibfield  {journal}
  {\bibinfo  {journal} {Phys. Rev. Lett.}\ }\textbf {\bibinfo {volume} {113}},\
  \bibinfo {pages} {156402} (\bibinfo {year} {2014})}\BibitemShut {NoStop}%
\bibitem [{\citenamefont {Liu}(2016)}]{liu2016topological}%
  \BibitemOpen
  \bibfield  {author} {\bibinfo {author} {\bibfnamefont {D.-P.}\ \bibnamefont
  {Liu}},\ }\bibfield  {title} {\bibinfo {title} {Topological phase boundary in
  a generalized kitaev model},\ }\href@noop {} {\bibfield  {journal} {\bibinfo
  {journal} {Chinese Physics B}\ }\textbf {\bibinfo {volume} {25}},\ \bibinfo
  {pages} {057101} (\bibinfo {year} {2016})}\BibitemShut {NoStop}%
\bibitem [{\citenamefont {Nehra}\ \emph
  {et~al.}(2018{\natexlab{b}})\citenamefont {Nehra}, \citenamefont {Bhakuni},
  \citenamefont {Gangadharaiah},\ and\ \citenamefont {Sharma}}]{nehra2018many}%
  \BibitemOpen
  \bibfield  {author} {\bibinfo {author} {\bibfnamefont {R.}~\bibnamefont
  {Nehra}}, \bibinfo {author} {\bibfnamefont {D.~S.}\ \bibnamefont {Bhakuni}},
  \bibinfo {author} {\bibfnamefont {S.}~\bibnamefont {Gangadharaiah}},\ and\
  \bibinfo {author} {\bibfnamefont {A.}~\bibnamefont {Sharma}},\ }\bibfield
  {title} {\bibinfo {title} {Many-body entanglement in a topological chiral
  ladder},\ }\href@noop {} {\bibfield  {journal} {\bibinfo  {journal} {Physical
  Review B}\ }\textbf {\bibinfo {volume} {98}},\ \bibinfo {pages} {045120}
  (\bibinfo {year} {2018}{\natexlab{b}})}\BibitemShut {NoStop}%
\bibitem [{\citenamefont {H{\"u}gel}\ and\ \citenamefont
  {Paredes}(2014)}]{hugel2014chiral}%
  \BibitemOpen
  \bibfield  {author} {\bibinfo {author} {\bibfnamefont {D.}~\bibnamefont
  {H{\"u}gel}}\ and\ \bibinfo {author} {\bibfnamefont {B.}~\bibnamefont
  {Paredes}},\ }\bibfield  {title} {\bibinfo {title} {Chiral ladders and the
  edges of quantum hall insulators},\ }\href@noop {} {\bibfield  {journal}
  {\bibinfo  {journal} {Physical Review A}\ }\textbf {\bibinfo {volume} {89}},\
  \bibinfo {pages} {023619} (\bibinfo {year} {2014})}\BibitemShut {NoStop}%
\bibitem [{\citenamefont {Roy}\ and\ \citenamefont
  {Sharma}(2018)}]{roy2018entanglement}%
  \BibitemOpen
  \bibfield  {author} {\bibinfo {author} {\bibfnamefont {N.}~\bibnamefont
  {Roy}}\ and\ \bibinfo {author} {\bibfnamefont {A.}~\bibnamefont {Sharma}},\
  }\bibfield  {title} {\bibinfo {title} {Entanglement contour perspective for
  “strong area-law violation” in a disordered long-range hopping model},\
  }\href@noop {} {\bibfield  {journal} {\bibinfo  {journal} {Physical Review
  B}\ }\textbf {\bibinfo {volume} {97}},\ \bibinfo {pages} {125116} (\bibinfo
  {year} {2018})}\BibitemShut {NoStop}%
\bibitem [{\citenamefont {Roy}\ and\ \citenamefont
  {Sharma}(2019)}]{roy2019study}%
  \BibitemOpen
  \bibfield  {author} {\bibinfo {author} {\bibfnamefont {N.}~\bibnamefont
  {Roy}}\ and\ \bibinfo {author} {\bibfnamefont {A.}~\bibnamefont {Sharma}},\
  }\bibfield  {title} {\bibinfo {title} {Study of counterintuitive transport
  properties in the aubry-andr$\backslash$'e-harper model via entanglement
  entropy and persistent current},\ }\href@noop {} {\bibfield  {journal}
  {\bibinfo  {journal} {arXiv preprint arXiv:1905.13255}\ } (\bibinfo {year}
  {2019})}\BibitemShut {NoStop}%
\bibitem [{\citenamefont {Dey}\ \emph {et~al.}(2019)\citenamefont {Dey},
  \citenamefont {Bhakuni}, \citenamefont {Agarwalla},\ and\ \citenamefont
  {Sharma}}]{dey2019quantum}%
  \BibitemOpen
  \bibfield  {author} {\bibinfo {author} {\bibfnamefont {A.}~\bibnamefont
  {Dey}}, \bibinfo {author} {\bibfnamefont {D.~S.}\ \bibnamefont {Bhakuni}},
  \bibinfo {author} {\bibfnamefont {B.~K.}\ \bibnamefont {Agarwalla}},\ and\
  \bibinfo {author} {\bibfnamefont {A.}~\bibnamefont {Sharma}},\ }\bibfield
  {title} {\bibinfo {title} {Quantum entanglement and transport in
  non-equilibrium interacting double-dot setup: The curious role of
  degeneracy},\ }\href@noop {} {\bibfield  {journal} {\bibinfo  {journal}
  {arXiv preprint arXiv:1902.00474}\ } (\bibinfo {year} {2019})}\BibitemShut
  {NoStop}%
\bibitem [{\citenamefont {Sharma}\ and\ \citenamefont
  {Rabani}(2015)}]{sharma2015landauer}%
  \BibitemOpen
  \bibfield  {author} {\bibinfo {author} {\bibfnamefont {A.}~\bibnamefont
  {Sharma}}\ and\ \bibinfo {author} {\bibfnamefont {E.}~\bibnamefont
  {Rabani}},\ }\bibfield  {title} {\bibinfo {title} {Landauer current and
  mutual information},\ }\href@noop {} {\bibfield  {journal} {\bibinfo
  {journal} {Physical Review B}\ }\textbf {\bibinfo {volume} {91}},\ \bibinfo
  {pages} {085121} (\bibinfo {year} {2015})}\BibitemShut {NoStop}%
\bibitem [{\citenamefont {Sable}\ \emph {et~al.}(2018)\citenamefont {Sable},
  \citenamefont {Bhakuni},\ and\ \citenamefont {Sharma}}]{sable2018landauer}%
  \BibitemOpen
  \bibfield  {author} {\bibinfo {author} {\bibfnamefont {H.~S.}\ \bibnamefont
  {Sable}}, \bibinfo {author} {\bibfnamefont {D.~S.}\ \bibnamefont {Bhakuni}},\
  and\ \bibinfo {author} {\bibfnamefont {A.}~\bibnamefont {Sharma}},\
  }\bibfield  {title} {\bibinfo {title} {Landauer current and mutual
  information in a bosonic quantum dot},\ }in\ \href@noop {} {\emph {\bibinfo
  {booktitle} {Journal of Physics: Conference Series}}},\ Vol.\ \bibinfo
  {volume} {964}\ (\bibinfo {organization} {IOP Publishing},\ \bibinfo {year}
  {2018})\ p.\ \bibinfo {pages} {012007}\BibitemShut {NoStop}%
\bibitem [{\citenamefont {Bhakuni}\ and\ \citenamefont
  {Sharma}(2018)}]{PhysRevB.98.045408}%
  \BibitemOpen
  \bibfield  {author} {\bibinfo {author} {\bibfnamefont {D.~S.}\ \bibnamefont
  {Bhakuni}}\ and\ \bibinfo {author} {\bibfnamefont {A.}~\bibnamefont
  {Sharma}},\ }\bibfield  {title} {\bibinfo {title} {Characteristic length
  scales from entanglement dynamics in electric-field-driven tight-binding
  chains},\ }\href {https://doi.org/10.1103/PhysRevB.98.045408} {\bibfield
  {journal} {\bibinfo  {journal} {Phys. Rev. B}\ }\textbf {\bibinfo {volume}
  {98}},\ \bibinfo {pages} {045408} (\bibinfo {year} {2018})}\BibitemShut
  {NoStop}%
\bibitem [{\citenamefont {Vitagliano}\ \emph {et~al.}(2010)\citenamefont
  {Vitagliano}, \citenamefont {Riera},\ and\ \citenamefont
  {Latorre}}]{vitagliano2010volume}%
  \BibitemOpen
  \bibfield  {author} {\bibinfo {author} {\bibfnamefont {G.}~\bibnamefont
  {Vitagliano}}, \bibinfo {author} {\bibfnamefont {A.}~\bibnamefont {Riera}},\
  and\ \bibinfo {author} {\bibfnamefont {J.~I.}\ \bibnamefont {Latorre}},\
  }\bibfield  {title} {\bibinfo {title} {Volume-law scaling for the
  entanglement entropy in spin-1/2 chains},\ }\href@noop {} {\bibfield
  {journal} {\bibinfo  {journal} {New Journal of Physics}\ }\textbf {\bibinfo
  {volume} {12}},\ \bibinfo {pages} {113049} (\bibinfo {year}
  {2010})}\BibitemShut {NoStop}%
\bibitem [{\citenamefont {Nehra}\ \emph
  {et~al.}(2019{\natexlab{b}})\citenamefont {Nehra}, \citenamefont {Sharma},\
  and\ \citenamefont {Soori}}]{nehra2019transport}%
  \BibitemOpen
  \bibfield  {author} {\bibinfo {author} {\bibfnamefont {R.}~\bibnamefont
  {Nehra}}, \bibinfo {author} {\bibfnamefont {A.}~\bibnamefont {Sharma}},\ and\
  \bibinfo {author} {\bibfnamefont {A.}~\bibnamefont {Soori}},\ }\bibfield
  {title} {\bibinfo {title} {Transport in a long-range kitaev ladder: role of
  majorana and subgap andreev states},\ }\href@noop {} {\bibfield  {journal}
  {\bibinfo  {journal} {arXiv preprint arXiv:1909.00565}\ } (\bibinfo {year}
  {2019}{\natexlab{b}})}\BibitemShut {NoStop}%
\bibitem [{\citenamefont {Xia}\ \emph {et~al.}(2018)\citenamefont {Xia},
  \citenamefont {Ramachandran}, \citenamefont {Xia}, \citenamefont {Li},
  \citenamefont {Liu}, \citenamefont {Tang}, \citenamefont {Hu}, \citenamefont
  {Song}, \citenamefont {Xu}, \citenamefont {Leykam} \emph
  {et~al.}}]{xia2018unconventional}%
  \BibitemOpen
  \bibfield  {author} {\bibinfo {author} {\bibfnamefont {S.}~\bibnamefont
  {Xia}}, \bibinfo {author} {\bibfnamefont {A.}~\bibnamefont {Ramachandran}},
  \bibinfo {author} {\bibfnamefont {S.}~\bibnamefont {Xia}}, \bibinfo {author}
  {\bibfnamefont {D.}~\bibnamefont {Li}}, \bibinfo {author} {\bibfnamefont
  {X.}~\bibnamefont {Liu}}, \bibinfo {author} {\bibfnamefont {L.}~\bibnamefont
  {Tang}}, \bibinfo {author} {\bibfnamefont {Y.}~\bibnamefont {Hu}}, \bibinfo
  {author} {\bibfnamefont {D.}~\bibnamefont {Song}}, \bibinfo {author}
  {\bibfnamefont {J.}~\bibnamefont {Xu}}, \bibinfo {author} {\bibfnamefont
  {D.}~\bibnamefont {Leykam}}, \emph {et~al.},\ }\bibfield  {title} {\bibinfo
  {title} {Unconventional flatband line states in photonic lieb lattices},\
  }\href@noop {} {\bibfield  {journal} {\bibinfo  {journal} {Physical review
  letters}\ }\textbf {\bibinfo {volume} {121}},\ \bibinfo {pages} {263902}
  (\bibinfo {year} {2018})}\BibitemShut {NoStop}%
\bibitem [{\citenamefont {Toikka}\ and\ \citenamefont
  {Andreanov}(2018)}]{toikka2018necessary}%
  \BibitemOpen
  \bibfield  {author} {\bibinfo {author} {\bibfnamefont {L.}~\bibnamefont
  {Toikka}}\ and\ \bibinfo {author} {\bibfnamefont {A.}~\bibnamefont
  {Andreanov}},\ }\bibfield  {title} {\bibinfo {title} {Necessary and
  sufficient conditions for flat bands in m-dimensional n-band lattices with
  complex-valued nearest-neighbour hopping},\ }\href@noop {} {\bibfield
  {journal} {\bibinfo  {journal} {Journal of Physics A: Mathematical and
  Theoretical}\ }\textbf {\bibinfo {volume} {52}},\ \bibinfo {pages} {02LT04}
  (\bibinfo {year} {2018})}\BibitemShut {NoStop}%
\bibitem [{\citenamefont {Alecce}\ and\ \citenamefont
  {Dell'Anna}(2017)}]{PhysRevB.95.195160}%
  \BibitemOpen
  \bibfield  {author} {\bibinfo {author} {\bibfnamefont {A.}~\bibnamefont
  {Alecce}}\ and\ \bibinfo {author} {\bibfnamefont {L.}~\bibnamefont
  {Dell'Anna}},\ }\bibfield  {title} {\bibinfo {title} {Extended kitaev chain
  with longer-range hopping and pairing},\ }\href
  {https://doi.org/10.1103/PhysRevB.95.195160} {\bibfield  {journal} {\bibinfo
  {journal} {Phys. Rev. B}\ }\textbf {\bibinfo {volume} {95}},\ \bibinfo
  {pages} {195160} (\bibinfo {year} {2017})}\BibitemShut {NoStop}%
\bibitem [{\citenamefont {Kitaev}(2009)}]{kitaev2009periodic}%
  \BibitemOpen
  \bibfield  {author} {\bibinfo {author} {\bibfnamefont {A.}~\bibnamefont
  {Kitaev}},\ }\bibfield  {title} {\bibinfo {title} {Periodic table for
  topological insulators and superconductors},\ }in\ \href@noop {} {\emph
  {\bibinfo {booktitle} {AIP conference proceedings}}},\ Vol.\ \bibinfo
  {volume} {1134}\ (\bibinfo {organization} {AIP},\ \bibinfo {year} {2009})\
  pp.\ \bibinfo {pages} {22--30}\BibitemShut {NoStop}%
\bibitem [{\citenamefont {Zhou}\ and\ \citenamefont
  {Lee}(2019)}]{PhysRevB.99.235112}%
  \BibitemOpen
  \bibfield  {author} {\bibinfo {author} {\bibfnamefont {H.}~\bibnamefont
  {Zhou}}\ and\ \bibinfo {author} {\bibfnamefont {J.~Y.}\ \bibnamefont {Lee}},\
  }\bibfield  {title} {\bibinfo {title} {Periodic table for topological bands
  with non-hermitian symmetries},\ }\href
  {https://doi.org/10.1103/PhysRevB.99.235112} {\bibfield  {journal} {\bibinfo
  {journal} {Phys. Rev. B}\ }\textbf {\bibinfo {volume} {99}},\ \bibinfo
  {pages} {235112} (\bibinfo {year} {2019})}\BibitemShut {NoStop}%
\bibitem [{\citenamefont {Wu}(2012{\natexlab{b}})}]{WU20123530}%
  \BibitemOpen
  \bibfield  {author} {\bibinfo {author} {\bibfnamefont {N.}~\bibnamefont
  {Wu}},\ }\bibfield  {title} {\bibinfo {title} {Topological phases of the
  two-leg kitaev ladder},\ }\href
  {https://doi.org/https://doi.org/10.1016/j.physleta.2012.10.016} {\bibfield
  {journal} {\bibinfo  {journal} {Physics Letters A}\ }\textbf {\bibinfo
  {volume} {376}},\ \bibinfo {pages} {3530 } (\bibinfo {year}
  {2012}{\natexlab{b}})}\BibitemShut {NoStop}%
\bibitem [{\citenamefont {Peschel}\ and\ \citenamefont
  {Eisler}(2009)}]{Peschel_2009}%
  \BibitemOpen
  \bibfield  {author} {\bibinfo {author} {\bibfnamefont {I.}~\bibnamefont
  {Peschel}}\ and\ \bibinfo {author} {\bibfnamefont {V.}~\bibnamefont
  {Eisler}},\ }\bibfield  {title} {\bibinfo {title} {Reduced density matrices
  and entanglement entropy in free lattice models},\ }\href
  {https://doi.org/10.1088/1751-8113/42/50/504003} {\bibfield  {journal}
  {\bibinfo  {journal} {Journal of Physics A: Mathematical and Theoretical}\
  }\textbf {\bibinfo {volume} {42}},\ \bibinfo {pages} {504003} (\bibinfo
  {year} {2009})}\BibitemShut {NoStop}%
\bibitem [{\citenamefont {Peschel}(2003)}]{peschel2003calculation}%
  \BibitemOpen
  \bibfield  {author} {\bibinfo {author} {\bibfnamefont {I.}~\bibnamefont
  {Peschel}},\ }\bibfield  {title} {\bibinfo {title} {Calculation of reduced
  density matrices from correlation functions},\ }\href@noop {} {\bibfield
  {journal} {\bibinfo  {journal} {Journal of Physics A: Mathematical and
  General}\ }\textbf {\bibinfo {volume} {36}},\ \bibinfo {pages} {L205}
  (\bibinfo {year} {2003})}\BibitemShut {NoStop}%
\bibitem [{\citenamefont {Peschel}(2012)}]{peschel2012special}%
  \BibitemOpen
  \bibfield  {author} {\bibinfo {author} {\bibfnamefont {I.}~\bibnamefont
  {Peschel}},\ }\bibfield  {title} {\bibinfo {title} {Special review:
  entanglement in solvable many-particle models},\ }\href@noop {} {\bibfield
  {journal} {\bibinfo  {journal} {Brazilian Journal of Physics}\ }\textbf
  {\bibinfo {volume} {42}},\ \bibinfo {pages} {267} (\bibinfo {year}
  {2012})}\BibitemShut {NoStop}%
\bibitem [{\citenamefont {Lieb}\ \emph {et~al.}(1961)\citenamefont {Lieb},
  \citenamefont {Schultz},\ and\ \citenamefont {Mattis}}]{lieb1961two}%
  \BibitemOpen
  \bibfield  {author} {\bibinfo {author} {\bibfnamefont {E.}~\bibnamefont
  {Lieb}}, \bibinfo {author} {\bibfnamefont {T.}~\bibnamefont {Schultz}},\ and\
  \bibinfo {author} {\bibfnamefont {D.}~\bibnamefont {Mattis}},\ }\bibfield
  {title} {\bibinfo {title} {Two soluble models of an antiferromagnetic
  chain},\ }\href@noop {} {\bibfield  {journal} {\bibinfo  {journal} {Annals of
  Physics}\ }\textbf {\bibinfo {volume} {16}},\ \bibinfo {pages} {407}
  (\bibinfo {year} {1961})}\BibitemShut {NoStop}%
\bibitem [{\citenamefont {Mahyaeh}\ and\ \citenamefont
  {Ardonne}(2018)}]{Mahyaeh_2018}%
  \BibitemOpen
  \bibfield  {author} {\bibinfo {author} {\bibfnamefont {I.}~\bibnamefont
  {Mahyaeh}}\ and\ \bibinfo {author} {\bibfnamefont {E.}~\bibnamefont
  {Ardonne}},\ }\bibfield  {title} {\bibinfo {title} {Zero modes of the kitaev
  chain with phase-gradients and longer range couplings},\ }\href
  {https://doi.org/10.1088/2399-6528/aab7e5} {\bibfield  {journal} {\bibinfo
  {journal} {Journal of Physics Communications}\ }\textbf {\bibinfo {volume}
  {2}},\ \bibinfo {pages} {045010} (\bibinfo {year} {2018})}\BibitemShut
  {NoStop}%
\bibitem [{\citenamefont {Vijay}\ and\ \citenamefont
  {Fu}(2015)}]{PhysRevB.91.220101}%
  \BibitemOpen
  \bibfield  {author} {\bibinfo {author} {\bibfnamefont {S.}~\bibnamefont
  {Vijay}}\ and\ \bibinfo {author} {\bibfnamefont {L.}~\bibnamefont {Fu}},\
  }\bibfield  {title} {\bibinfo {title} {Entanglement spectrum of a random
  partition: Connection with the localization transition},\ }\href
  {https://doi.org/10.1103/PhysRevB.91.220101} {\bibfield  {journal} {\bibinfo
  {journal} {Phys. Rev. B}\ }\textbf {\bibinfo {volume} {91}},\ \bibinfo
  {pages} {220101} (\bibinfo {year} {2015})}\BibitemShut {NoStop}%
\bibitem [{\citenamefont {Borchmann}\ and\ \citenamefont
  {Pereg-Barnea}(2017)}]{borchmann2017analytic}%
  \BibitemOpen
  \bibfield  {author} {\bibinfo {author} {\bibfnamefont {J.}~\bibnamefont
  {Borchmann}}\ and\ \bibinfo {author} {\bibfnamefont {T.}~\bibnamefont
  {Pereg-Barnea}},\ }\bibfield  {title} {\bibinfo {title} {Analytic expression
  for the entanglement entropy of a two-dimensional topological
  superconductor},\ }\href@noop {} {\bibfield  {journal} {\bibinfo  {journal}
  {Physical Review B}\ }\textbf {\bibinfo {volume} {95}},\ \bibinfo {pages}
  {075152} (\bibinfo {year} {2017})}\BibitemShut {NoStop}%
\bibitem [{\citenamefont {Yang}(2006)}]{YANG2006249}%
  \BibitemOpen
  \bibfield  {author} {\bibinfo {author} {\bibfnamefont {D.}~\bibnamefont
  {Yang}},\ }\bibfield  {title} {\bibinfo {title} {A simple proof of monogamy
  of entanglement},\ }\href
  {https://doi.org/https://doi.org/10.1016/j.physleta.2006.08.027} {\bibfield
  {journal} {\bibinfo  {journal} {Physics Letters A}\ }\textbf {\bibinfo
  {volume} {360}},\ \bibinfo {pages} {249 } (\bibinfo {year}
  {2006})}\BibitemShut {NoStop}%
\bibitem [{\citenamefont {Deng}\ \emph {et~al.}(2014)\citenamefont {Deng},
  \citenamefont {Ortiz}, \citenamefont {Poudel},\ and\ \citenamefont
  {Viola}}]{PhysRevB.89.140507}%
  \BibitemOpen
  \bibfield  {author} {\bibinfo {author} {\bibfnamefont {S.}~\bibnamefont
  {Deng}}, \bibinfo {author} {\bibfnamefont {G.}~\bibnamefont {Ortiz}},
  \bibinfo {author} {\bibfnamefont {A.}~\bibnamefont {Poudel}},\ and\ \bibinfo
  {author} {\bibfnamefont {L.}~\bibnamefont {Viola}},\ }\bibfield  {title}
  {\bibinfo {title} {Majorana flat bands in $s$-wave gapless topological
  superconductors},\ }\href {https://doi.org/10.1103/PhysRevB.89.140507}
  {\bibfield  {journal} {\bibinfo  {journal} {Phys. Rev. B}\ }\textbf {\bibinfo
  {volume} {89}},\ \bibinfo {pages} {140507} (\bibinfo {year}
  {2014})}\BibitemShut {NoStop}%
\bibitem [{\citenamefont {Kumar}\ \emph {et~al.}(2019)\citenamefont {Kumar},
  \citenamefont {T\"orm\"a},\ and\ \citenamefont
  {Vanhala}}]{PhysRevB.100.125141}%
  \BibitemOpen
  \bibfield  {author} {\bibinfo {author} {\bibfnamefont {P.}~\bibnamefont
  {Kumar}}, \bibinfo {author} {\bibfnamefont {P.}~\bibnamefont {T\"orm\"a}},\
  and\ \bibinfo {author} {\bibfnamefont {T.~I.}\ \bibnamefont {Vanhala}},\
  }\bibfield  {title} {\bibinfo {title} {Magnetization, $d$-wave
  superconductivity, and non-fermi-liquid behavior in a crossover from
  dispersive to flat bands},\ }\href
  {https://doi.org/10.1103/PhysRevB.100.125141} {\bibfield  {journal} {\bibinfo
   {journal} {Phys. Rev. B}\ }\textbf {\bibinfo {volume} {100}},\ \bibinfo
  {pages} {125141} (\bibinfo {year} {2019})}\BibitemShut {NoStop}%
\end{thebibliography}%
\section*{Appendix 1: Flat bands in Kitaev chain}

The Kitaev ladder discussed in the present article has been shown to possess two or four flat bands according to the parameter set chosen for the Hamiltonian. Further, the Kitaev ladder has been mapped on to an interlinked cross-stitch lattice and the eigenstates corresponding to flat bands have been shown as CLS. In this Appendix, we show that setting the rungs hopping $t'=0$ separates the Kitaev ladder into two Kitaev chains and the latter can be mapped into a well-known flat band lattice known as cross-stitch lattice. 

When $t'$ is set to zero, the tight-binding Hamiltonian for Kitaev ladder leads to that for a Kitaev chain and can be represented as:
\begin{equation}
\begin{split}
H=&-t\displaystyle\sum_{n}c_{n+1}^{\dagger}c_{n}-\mu\displaystyle\sum_{n}c_{n}^{\dagger}c_{n}\\&-\Delta\displaystyle\sum_{n}c_{n+1}^{\dagger}c_{n}^{\dagger}+H~.~c~,
\end{split}\label{Ham_rs_1}
\end{equation}
where $t$ is the intra-leg hopping amplitude,
$c^{\dagger}_{n}$($c_{n}$) are creation (annihilation)
operators on the $n^{th}$ site of the chain.
Through a Bogoliubov transformation~\citep{nehra2019enhancement,nehra2019transport}, the overall Hamiltonian in $k-$space is given as
\begin{align}\small
\mathcal{H}(k)=\begin{bmatrix}
-\epsilon_{\mu ,k} &t_{\Delta} \\
t^*_{\Delta} & \epsilon_{\mu ,k} \\
\end{bmatrix} \label{ham_k_1}
\end{align}
with $\epsilon_{\mu ,k} = (2t\cos{k}+\mu)$ and $t_{\Delta}=-2i\Delta\sin{k}$.  The dispersion relation corresponding to the Kitaev chain can be obtained as
\begin{equation}
 E(k)=\pm\sqrt{\epsilon_{\mu ,k}^2+\tau_{\Delta,k}}
 \label{disp_chain_1}
\end{equation} 
with $\epsilon_{\mu ,k} = (2t\cos{k}+\mu)$ and $\tau_{\Delta,k} = 4{\Delta}^2 {\sin}^2 k$.
The band structure consists of two bands which reflect the particle-hole symmetry of the Hamiltonian. 

\begin{figure}
\begin{center}
\includegraphics[]{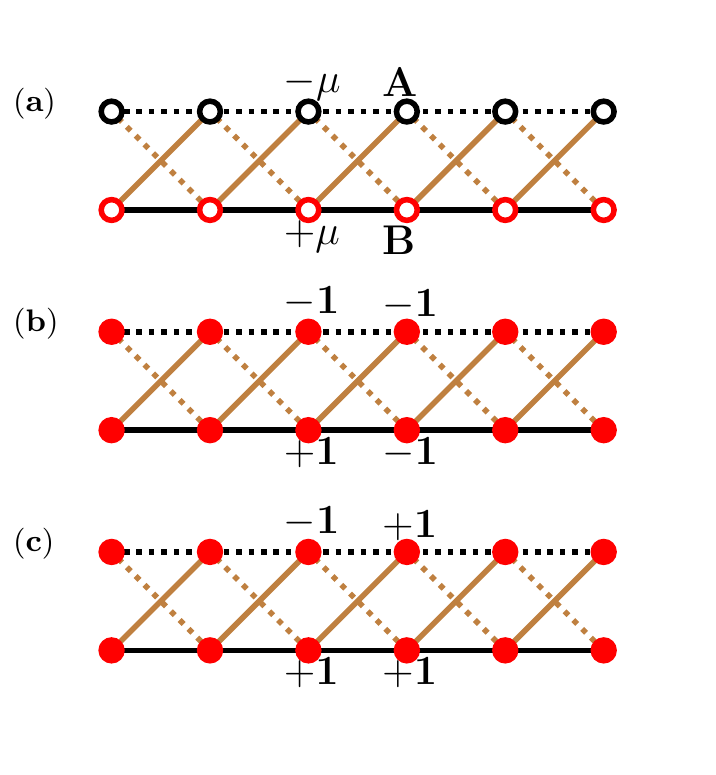}
\caption{(a) Schematic representation of the cross-stitch chain. The black (red) open circles represent negative (positive) onsite energies at sites and red dots represents $\mu = 0$. The dotted (regular) lines represent negative (positive) hopping. (b)  and (c) The eigenstates corresponding to $E = -2$ and $E=+2$ respectively for $\mu = 0,~t = \Delta = 1$.  }
\label{k_chain_1}
\end{center}
\end{figure}

To understand
the flat band properties of the Kitaev chain better, the Hamiltonian
can be rearranged as
\begin{equation}
H = 
\left(
\begin{array}{cccc}
 -\mu-e^{-i k} t-e^{i k} t & \Delta e^{-i k}-\Delta e^{i k} \\
 \Delta e^{i k}-\Delta e^{-i k} & \mu+e^{-i k} t+e^{i k} t \\
\end{array}
\right).
\label{app_Ham_1}
\end{equation}
The resulting Hamiltonian can be mapped onto a cross-stitch chain as shown in Fig.~\ref{k_chain_1}(a). The cross-stitch chain consists of two sites per unit cell represented by $A$ and $B$ in the figure.  For $\mu = 0,~t = \Delta = 1$, the Kitaev chain and hence the cross-stitch chain supports two flat bands at $E = \pm 2$. The CLS corresponding to both flat bands in the cross-stitch chain are shown in Fig.~\ref{k_chain_1}(b) and (c). The CLS for both bands reside on two unit cells (i.e., four sites) with strictly zero amplitudes in the rest of the chain.

\end{document}